\documentclass[preprint,notoc]{JHEP3}
\usepackage{epsfig, amssymb, amsmath}
\usepackage{epstopdf}
\usepackage{float}

\def\eslt{\not\!\!{E_T}}

\def\to{\rightarrow}

\def\bi{\begin{itemize}}
 \def\ei{\end{itemize}}

\def\c1p{C1^\prime}

\def\tu{\tilde u}

\def\tst{\tilde t}

\def\tg{\tilde g}
\def\tnu{\tilde\nu}

\def\tq{\tilde q}

\def\tw{\widetilde W}
\def\tz{\widetilde Z}

\def\anu{A_{\tilde\nu}}

\def\lsp{\tilde{\nu}_1}

\def\agt{\gtrsim}
\def\be{\begin{equation}}  
\def\ee{\end{equation}}  
\def\bea{\begin{eqnarray}}  
\def\eea{\end{eqnarray}}  

\def\pT{\mathbf{p_T}}

\title{Light Sneutrino Dark Matter at the LHC}

\author{Genevi\`eve B\'elanger$^{a}$, Sabine Kraml$^{b}$, Andre Lessa$^{c}$\\
$^a$LAPTH, Univ. de Savoie, CNRS, B.P. 110, F-74941, Annecy-le-Vieux, France\\
$^b$Laboratoire de Physique Subatomique et de Cosmologie, UJF Grenoble 1, 
CNRS/IN2P3, INPG, 53 Avenue des Martyrs, F-38026 Grenoble, France\\
$^c$Dept.\ of Physics and Astronomy, University of Oklahoma, Norman, OK 73019, USA\\

E-mail: \email{belanger@lapp.in2p3.fr}, \email{sabine.kraml@lpsc.in2p3.fr}, \email{lessa@nhn.ou.edu}}

\abstract{In supersymmetric (SUSY) models with Dirac neutrino masses,  
a weak-scale trilinear $A_{\tilde\nu}$ term that is not proportional 
to the small neutrino Yukawa couplings can induce a sizable 
mixing between left and right-handed sneutrinos. The lighter 
sneutrino mass eigenstate can hence become the lightest SUSY 
particle (LSP) and a viable dark matter candidate. 
In particular, it can be an excellent candidate for light dark matter 
with mass below $\sim$10~GeV. Such a light mixed sneutrino LSP 
has a dramatic effect on SUSY signatures at the LHC, as charginos 
decay dominantly into the light sneutrino, $\lsp$, plus a charged lepton,  
and neutralinos decay invisibly into $\lsp\nu$. 
We perform a detailed study of the LHC potential 
to resolve the light sneutrino dark matter scenario 
by means of three representative benchmark points with 
different gluino and squark mass hierarchies. 
We study in particular the determination of the $\lsp$ 
mass from cascade decays involving charginos, using the $m_{T2}$ variable.
Moreover, we address 
measurements of additional invisible sparticles, in our case 
$\tz_{1,2}$, and the question of discrimination against the 
MSSM. 
}


\begin{document}

\section{Introduction} \label{sec:intro}

There is abundant and compelling evidence that the bulk of matter in the universe is 
made of massive, electrically neutral particles: dark matter (DM)~\cite{Bertone:2010zz}. 
While the density of DM has been precisely determined, the identity of the DM particle 
(or particles) is a complete mystery. 
Remarkably, many extensions of the Standard Model (SM) of electroweak and strong 
interactions---designed to address theoretical issues related to the breaking of the 
electroweak symmetry---require the introduction of new particles, some of which are 
excellent DM candidates. This is most notably the case in R-parity conserving 
supersymmetry (SUSY)~\cite{Jungman:1995df}. 

If the origin of DM is indeed new physics beyond the SM, in particular SUSY, there are high hopes that 
DM will be produced in abundance at the LHC, through cascade decays of the new 
matter particles produced in the pp collisions~\cite{Baer:2008uu}.  
The ambitious goal will then be to determine the properties of the DM candidate and reconstruct 
its thermal relic abundance from collider data. On the one hand this may then be used further 
to make testable predictions for direct and indirect DM searches. On the other hand, if the
DM mass determined at colliders and non-accelerator experiments agrees, it may be used to 
constrain the cosmological model. This would enormously enhance the 
interplay between particle physics, astrophysics and cosmology. 
The LHC phenomenology of DM-motivated SUSY scenarios has hence been 
discussed in great detail in the literature~\cite{Jungman:1995df,Baer:2008uu}.

Another experimental evidence that the SM misses something fundamental 
is the observation of neutrino oscillations~\cite{Bilenky:1998dt}.
This can be resolved by including right-handed neutrinos in the model. 
Current observations, however, do not allow to establish the Majorana or 
Dirac nature of neutrinos. While the smallness of the neutrino mass can be 
naturally explained by introducing Majorana mass terms and making use of 
the see-saw mechanism, Dirac masses for neutrinos with very small Yukawa 
couplings are a viable and interesting alternative. 
In supersymmetric theories,  one may naturally obtain very light Dirac neutrino 
masses from F-term SUSY breaking~\cite{ArkaniHamed:2000bq,Borzumati:2000mc}. 
In addition to providing an explanation for neutrino masses, this class of SUSY 
models offers an interesting alternative to the conventional neutralino DM candidate: the sneutrino. 

The crucial point is that in these models one can have a weak-scale trilinear 
$A_{\tilde\nu}$ term that is not proportional to the small neutrino Yukawa couplings and can hence induce a large mixing between left-handed (LH) and right-handed (RH) 
sneutrinos even though the Yukawa couplings are extremely small.  
The lightest sneutrino can thus become the lightest SUSY particle (LSP) 
and a viable thermal DM candidate.   

Note that the mainly RH sneutrino LSP is not sterile but couples to SM gauge and Higgs 
bosons through the mixing with its LH partner. Sufficient mixing provides efficient 
annihilation so that the mixed sneutrino can be a viable thermal DM candidate 
with a relic density of $\Omega h^2 \simeq 0.11$ as 
extracted from cosmological observations~\cite{Komatsu:2010fb,Jarosik:2010iu}. 
On the other hand the amount of mixing is constrained by limits on the 
spin-independent scattering cross-section, for which the LH sneutrino component 
receives an important contribution from Z exchange; this cross-section is suppressed 
by the sneutrino mixing angle. 
Because of the gauge and Higgs interactions, the presence of the 
mixed sneutrino can also significantly impact Higgs and SUSY signatures 
at the LHC~\cite{ArkaniHamed:2000bq}.

In \cite{Belanger:2010cd}, some of us investigated the case of mixed sneutrinos as 
thermal DM with special emphasis on the mass range below $\sim$10~GeV. 
We examined the viable parameter space and discussed implications for direct 
and indirect dark matter searches, as well as consequences for collider phenomenology.
Regarding the latter, we found that the SUSY signatures greatly differ from the expectations in the conventional Minimal Supersymmetric 
Standard Model (MSSM) with a $\tz_1$ LSP:
while squarks and gluinos have the usual cascade decays through charginos and 
neutralinos, with the same branching ratios as in the corresponding MSSM case, 
the charginos and neutralinos decay further into the $\lsp$ LSP. 
In particular,
\begin{equation}
    \tz_{1,2}\to \nu\,\lsp, \qquad \tw_1\to l^\pm\,\lsp 
\end{equation}
with practically 100\% branching ratio over most of the parameter space. 
At the LHC, the typical cascade decays therefore are 
$\tilde q_R^{}\to q\tz_1\to q\nu\lsp$, 
$\tilde q_L^{}\to q\tz_2\to q\nu\lsp$ and 
$\tilde q_L^{}\to q\tw_1\to q'l^\pm\lsp$, 
all giving different amount of missing transverse energy, $\eslt$.  
Moreover, gluino-pair production followed by decays into $qq'\tw_1$ 
through either on- or off-shell squarks leads to  
same-sign (SS) and opposite-sign (OS) dilepton events 
with equal probability.
Besides, the light Higgs decays  invisibly into a pair 
of LSPs, $h^0\to\lsp\lsp$. 

In this paper, we now perform a detailed study of the LHC potential 
to resolve the light sneutrino DM scenario. 
This includes in particular the determination of the DM mass 
from $\tilde g\to qq'\tw_1\to q'l^\pm\tilde\nu_{1}$ and/or 
$\tilde q\to q'\tw_1\to q'l^\pm\tilde\nu_{1}$ events. 
To this end we rely on the subsystem $m_{T2}$ method~\cite{Mt2sub}.  
Moreover, we address the question of 
measuring the masses of additional invisible sparticles. 
The case of a $\sim 100$~GeV mixed sneutrino LSP 
was studied in \cite{Thomas:2007bu}.

The paper is organized as follows. First, 
in Section~\ref{sec:framework}, we briefly recall the main features of 
the mixed sneutrino model. In Section~\ref{sec:bench} we present 
three benchmark points and their characteristic signatures. 
This sets the framework of the analysis.  We then go on in 
Section~\ref{sec:Disc} to study the discovery potential at 
the LHC with 7~TeV center-of-mass energy. 
Measurements at the 14~TeV LHC are discussed in detail in Section~\ref{sec:Masses}.
A summary and conclusions are given in Section~\ref{sec:conclude}.

\section{Mixed sneutrinos} \label{sec:framework}

The framework for our study is the model of \cite{ArkaniHamed:2000bq} with only Dirac masses 
for sneutrinos. In this case, the usual MSSM soft-breaking terms are extended by
\begin{equation}
  \Delta {\cal L}_{\rm soft} = m^2_{\tilde N_i}  |\tilde N_i |^2 +  
                                            A_{\tilde\nu_i} \tilde L_i \tilde N_i H_u + {\rm h.c.} \,,
\end{equation}
where ${m}^2_{\tilde{N}}$ and $A_{\tilde\nu}$ are weak-scale soft terms, which we assume to 
be flavor-diagonal. Note that the lepton-number violating bilinear term, which appears 
in case of Majorana neutrino masses, is absent. 
Neglecting the tiny Dirac masses, the $2\times2$ sneutrino mass matrix for one generation is 
given by 
\begin{equation}
  m^2_{\tilde\nu} =
  \left( \begin{array}{cc}
   {m}^2_{\widetilde{L}} +\frac{1}{2} m^2_Z \cos 2\beta \quad &  \frac{1}{\sqrt{2}} A_{\tilde\nu}\, v \sin\beta\\
   \frac{1}{\sqrt{2}}    A_{\tilde\nu}\, v \sin\beta&  {m}^2_{\widetilde{N}}
  \end{array}\right) \,.
\label{eq:sneutrino_tree}
\end{equation}
Here ${m}^2_{\tilde{L}}$ is the SU(2) slepton soft term, $v^2=v_1^2+v_2^2=(246\;{\rm GeV})^2$ 
with $v_{1,2}$ the Higgs vacuum expectation values, and $\tan\beta=v_2/v_1$.  
The main feature of this model is that the ${m}^2_{\widetilde{L}}$, ${m}^2_{\widetilde{N}}$
and $A_{\tilde\nu}$ are all of the order of the weak scale, and 
$A_{\tilde\nu}$ does not suffer any suppression from Yukawa couplings. 
In the following, we will always assume $m_{\widetilde N}<m_{\widetilde L}$ so that the lighter mass 
eigenstate, $\tilde\nu_1$, is mostly a $\tilde\nu_R$. 
This is in fact well motivated from renormalization group evolution, 
since for the gauge-singlet $\tilde\nu_R$ the running at 1~loop is driven exclusively by the 
$A_{\tilde\nu}$ term: 
\begin{equation}
  \frac{dm_{\widetilde N}^2}{dt} = \frac{2}{16\pi^2}\anu^2 \,,  
\end{equation}
while
\begin{equation}
  \frac{dm_{\widetilde L}^2}{dt} = -\frac{3}{16\pi^2}g_2^2M_2^2 
                                                    -\frac{3}{80\pi^2}g_Y^2M_1^2 
                                                    +\frac{1}{16\pi^2}\anu^2 \,.
\end{equation}
 
A large $A_{\tilde\nu}$ term in the sneutrino mass matrix will induce a significant 
mixing between the RH and LH states, 
\begin{equation}
  \left(\begin{array}{c}
    \tnu_{1}\\
    \tnu_{2}
  \end{array}\right) = 
  \left(\begin{array}{lr}
    \cos\theta_{\tnu}\, & -\sin\theta_{\tnu}\\
    \sin\theta_{\tnu} & \cos\theta_{\tnu}
  \end{array}\right) 
  \left(\begin{array}{c}
    \tnu_{R}\\
    \tnu_{L}
  \end{array}\right) ,
  \quad
  \sin2\theta_{\tnu} = 
     \frac{\sqrt{2} A_{\tnu} v \sin\beta}{m_{\tnu_2}^2 - m_{\tnu_1}^2}\,,
\end{equation}
leading to mass eigenvalues  
\begin{equation}
  m_{\tnu_{1,2}}^2 = 
  \frac{1}{2} \left(m_{+}^2 \mp \sqrt{m_{-}^4 + 2 A_{\tnu}^2 v^2 \sin^2\beta}\right) 
\end{equation}
where $m_{\pm}^2 \equiv m_{\tilde{L}}^2 \pm m_{\tilde{N}}^2 + m_Z^2/2$, 
and $m_{\tilde\nu_1} < m_{\tilde\nu_2}$ by convention.
%
Notice that a large value of $\anu$ can induce a large splitting between the two mass eigenstates 
even if ${m}^2_{\widetilde{L}}$ and ${m}^2_{\widetilde{N}}$ are of the same order, leading 
to scenarios where $m_{\tilde\nu_1} \ll m_{\tilde\nu_2},m_{\tilde{l}_L}$. In this way, 
$\tilde\nu_1$ can naturally be driven much below the neutralino masses.

Taking $m_{\tnu_{1}}$, $m_{\tnu_{2}}$ and $\theta_{\tnu}$ as input, the soft 
terms $m_{\widetilde N}$, $m_{\widetilde L}$ and $A_{\tnu}$ entering 
the sneutrino mass matrix Eq.~(\ref{eq:sneutrino_tree}) are fixed. 
This also fixes the corresponding LH charged slepton mass, 
$m_{\tilde l_L}^2= m_{\widetilde L}^2 + m_Z^2\cos 2\beta(\sin^2\theta_W-\frac{1}{2})$. 
For the RH one, $m_{\tilde l_R}^2= m_{\widetilde R}^2 - m_Z^2\cos 2\beta\sin^2\theta_W$, we assume $m_{\widetilde R}\equiv m_{\widetilde L}$ for simplicity.
We use an appropriately modified version of {\tt SuSpect}~\cite{Suspect} for the spectrum calculation, which includes in particular radiative corrections induced by the $A_{\tnu}$ term, as given in \cite{Belanger:2010cd}. 

A full scan of the relevant parameter space was done in \cite{Belanger:2010cd},
taking into account constraints from the Z invisible decay width, the Higgs 
and SUSY mass limits, as well as DM constraints from the relic abundance and 
direct and indirect DM searches. It was found that light mixed sneutrino DM 
consistent with all constraints populates the region
\begin{equation}
   1 \mbox{ GeV} \lesssim m_{\tnu_{1}} \lesssim 8 \mbox{ GeV} 
   \quad\mbox{and}\quad  0.1 \lesssim \sin\theta_{\tnu} \lesssim 0.4 \,.
\end{equation}
Moreover, over most of the parameter space, $m_{\tnu_2} \gtrsim 200$~GeV and 
$m_{\tilde{l}_{L,R}} > m_{\tz_1,\tw_1}$. 
For LHC physics this means that the final steps of the cascade
decays will be dominated by $\tz_{1,2} \to \nu+\lsp$ and 
$\tw_1 \to l+\lsp$. The remaining relevant parameters are the 
squark and gluino masses, which determine the production cross sections.

\section{Benchmark points and characteristic signatures}
\label{sec:bench}

In order to study the light mixed sneutrino DM (SNDM) scenario at the LHC,
we pick a parameter point from~\cite{Belanger:2010cd} 
with a $\tnu_1$ LSP of $7.6$~GeV as the DM candidate. 
The other sneutrino parameters are    
$m_{\tnu_2}=524$~GeV, $\sin\theta_{\tnu}=0.225$ and $\tan\beta=10$ 
($A_\nu=348$~GeV). The neutralino/chargino sector
is given by  $M_2=2M_1=221$~GeV and $\mu=800$~GeV. Together with 
$m_{\widetilde R}=m_{\widetilde L}=514$~GeV, this fixes the properties  
of the weakly interacting sparticles. 
Note we assume flavor degeneracy.

Based on this setup we define in Table~\ref{tab:bm} three benchmark 
points (SN1--SN3) with the same electroweak sector but different 
gluino--squark mass hierarchies. 
The squark and gluino production cross sections, computed using Pythia\cite{Pythia}, are given in Table~\ref{tab:xsects}, 
and the relevant decay branching ratios  are given in Table~\ref{tab:decays}. 

\begin{table}\centering
\begin{tabular}{lrrr}
\hline
 & SN1  & SN2 & SN3  \\
\hline
$m_{\tg}$   &  765 &  765 &  1000 \\
$m_{\tu_L}$ &  1521 &  775 &  700 \\
$m_{\tu_R}$ &  1521 &  776 &  700 \\
$m_{\tilde b_1}$ & 1514 & 766 & 689 \\
$m_{\tst_1}$ & 1441 & 675 & 584 \\
\hline
$m_{\tw_2}$ & 811 & 807 & 805 \\
$m_{\tnu_2}$ & 524 &  524 & 524 \\
$m_{\tilde e_{L,R}}$ & 516 & 516 & 516 \\
$m_{\tilde\tau_1}$ & 503 & 503 & 503 \\
$m_{\tw_1}$ & 227 & 228 & 228 \\
$m_{\tz_2}$ & 227 & 228 & 228 \\ 
$m_{\tz_1}$ & 109 &  109 & 109 \\ 
$m_{\tnu_1}$ & 7.6 &  7.6 & 7.6 \\
\hline
\end{tabular}
\caption{Masses in~GeV units for the SNDM benchmark points SN1--SN3.  
The spectrum is computed at the EW scale using a modified version of {\tt Suspect}~\cite{Suspect}. 
For comparison with the MSSM case, we consider points MSSM1--MSSM3 with the same masses  
as SN1--SN3 but the ${\tnu_1}$ removed from the spectrum.}
\label{tab:bm}
\end{table}

\begin{table}\centering
\begin{tabular}{l|rrr|rrr}
       & \multicolumn{3}{|c|}{7~TeV LHC} & \multicolumn{3}{c}{14~TeV LHC} \\
       &  $\tg\tg$ & $\tg\tq$ & $\tq\tq$ &  $\tg\tg$ & $\tg\tq$ & $\tq\tq$ \\ 
\hline
   SN1 & 0.03 & 0.008 & 0.0 & 1.1 & 0.5 & 0.05 \\
   SN2 & 0.02 & 0.2 & 0.2 & 1.0 & 4.3 & 2.3 \\
   SN3 & 0.002 & 0.08 & 0.35 & 0.2  & 2.0  & 3.2  \\
\hline
\end{tabular}
\caption{Cross sections in [pb] for gluino and squark production at 7~TeV and 14~TeV. }\label{tab:xsects}
\end{table}

\begin{table}\centering
\begin{tabular}{llrrr}
\hline
parent & daughters & SN1  & SN2 & SN3 \\
\hline \hline
$\tg \to$ & $q\,\tilde q_{L,R}$ & --\quad\ & --\quad\ & 71\%\\
          & $b\,\tilde b_{1,2}$ & --\quad\ & --\quad\ & 17\%\\
          & $t\,\tilde t_{1,2}$ & --\quad\ & --\quad\ & 12\%\\
          & $q\bar q'\,\tw_1$ & 52\% & 53\% & --\quad\ \\
          & $q\bar q\,\tz_2$  & 33\% & 33\%& --\quad\ \\ 
          & $q\bar q\,\tz_1$  & 15\% & 14\%& --\quad\ \\ 
\hline
$\tilde q_L\to$ & $q \tg$   & 72\% & 0.4\% &  --\quad\ \\
                & $q'\tw_1$ & 18\% & 66\% & 66\% \\ 
                & $q \tz_2$ & 9\% & 33\% & 33\% \\
\hline
$\tilde q_R\to$ & $q \tg$   & 93\% & 2\% & --\quad\ \\
                & $q \tz_1$ & 7\% &  98\% & 100\% \\
\hline
$\tst_1\to$ & $t \tg$   & 64\% & --\quad\  & --\quad\ \\
            & $b \tw_1$ &  9\%  & 58\% & 59\%\\
            & $t \tz_2$ &  4\%  & 24\% & 23\%\\
            & $t \tz_1$ &  3\%  & 17\% & 18\%\\
\hline
$\tw_1^\pm \to$ & $l^\pm\tnu_1$ & 99\% & 99\% & 99\% \\ 
$\tz_2 \to$ & $\nu\tnu_1$   & 100\% & 100\% & 100\% \\ 
$\tz_1 \to$ & $\nu\tnu_1$   & 100\% & 100\% & 100\% \\ 
\hline
\end{tabular}
\caption{Most important decay channels for points SN1--SN3.}\label{tab:decays}
\end{table}

Employing GUT relations for gaugino-mass parameters, 
point SN1 has a gluino mass of $m_{\tg}=765$~GeV,  
while squark soft terms are set to 1.5~TeV, resulting in  
$m_{\tilde q}\simeq 2m_{\tg}$. 
Point SN1 is therefore characterized by heavy quarks and a light gluino, 
which decays through 3-body final states into charginos and neutralinos,
which will then decay exclusively to
$l^{\pm} \tnu_1$ and $\nu \tnu_1$, respectively. We hence expect dominantly 
gluino-pair production leading to 4 jets plus $\eslt$.
Moreover, about half of the events will have an isolated charged lepton, 
and about 25\% of the events will have two leptons with uncorrelated flavor 
(assuming three roughly degenerate light sneutrinos, 2/3 of the leptons 
are electrons or muons and 1/3 are taus). 
Since the gluino is a Majorana particle, same-sign and opposite-sign 
dileptons should have equal rates. However, same-sign dileptons 
have less SM background. Promising signatures to look for are hence 
\begin{quote}
(a) 4 jets, 0 leptons, large $\eslt$; \\
(b) 4 jets, same-sign dileptons, moderate $\eslt$; \\
(c) 4 jets, opposite-sign dileptons, moderate $\eslt$.
\end{quote}

Point SN2 has the same gluino mass as SN1 
but lighter squarks with $m_{\tilde q}\sim m_{\tg}$. 
Therefore SN2 has a much larger overall SUSY production, since the squark-pair and gluino-squark 
associated production give the main contributions to the total cross-section, as shown in Table~\ref{tab:xsects}.
The gluinos have the same decay modes as above, while the squarks decay dominantly through $q+\tw_1/\tz_{1,2}$.
 Events with only 2 or 3 jets are predominant and result from $\tq\tq$  
or $\tg\tq$ production, with the squark decaying into a neutralino or chargino 
(99\% of the $\tq_L$ and 98\% of the $\tq_R$ decays).

Finally, point SN3 is characterized by light squarks and heavy 
gluinos (achieved through non-universal gaugino masses). 
We hence expect dominantly squark-pair production followed by decays 
into quarks plus charginos or neutralinos, see Table~\ref{tab:decays}. 
These events have 2 hard jets plus $\eslt$, often accompanied by 1--2 
leptons from $\tq_L\to q\tw_1\to ql\tnu_1$. Again, events without leptons 
are expected to have larger $\eslt$ on average than events with leptons. 

It is also interesting to compare the phenomenology of the three 
SNDM benchmark points with the corresponding MSSM cases. To this 
end we consider points MSSM1--MSSM3 with the same masses  
as SN1--SN3 in Table~\ref{tab:bm} but with the ${\tnu_1}$ removed 
from the spectrum. The MSSM points thus have a neutralino LSP 
with a mass of 109~GeV. The $\tw_1$ decays exclusively into $\tz_1W$, 
while the $\tz_2$ decays into $\tz_1Z$ (12\%) or into $\tz_1h$ (88\%).\footnote{In 
the SNDM case, these decays are highly suppressed 
with respect to the decay into the sneutrino LSP. Furthermore,
for SN1--SN3, $h^0\to \lsp\lsp$ with practically 100\% branching ratio.} 
The larger rate of multilepton events in the SNDM scenario is a clear difference 
to the MSSM case. On the other hand one can expect MSSM events to have 
higher jet multiplicity.

\FIGURE[t]{
\includegraphics[width=7cm]{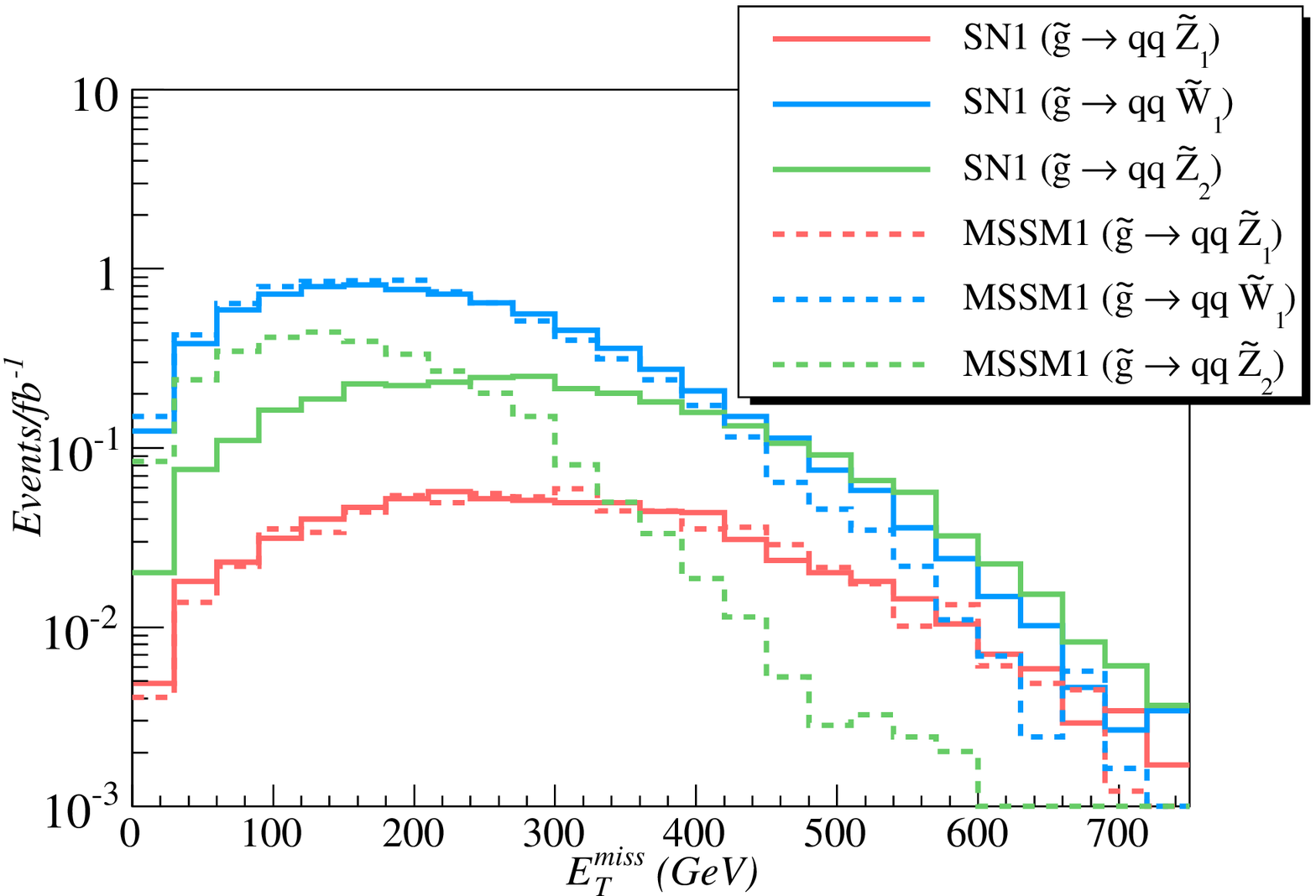}
\includegraphics[width=7cm]{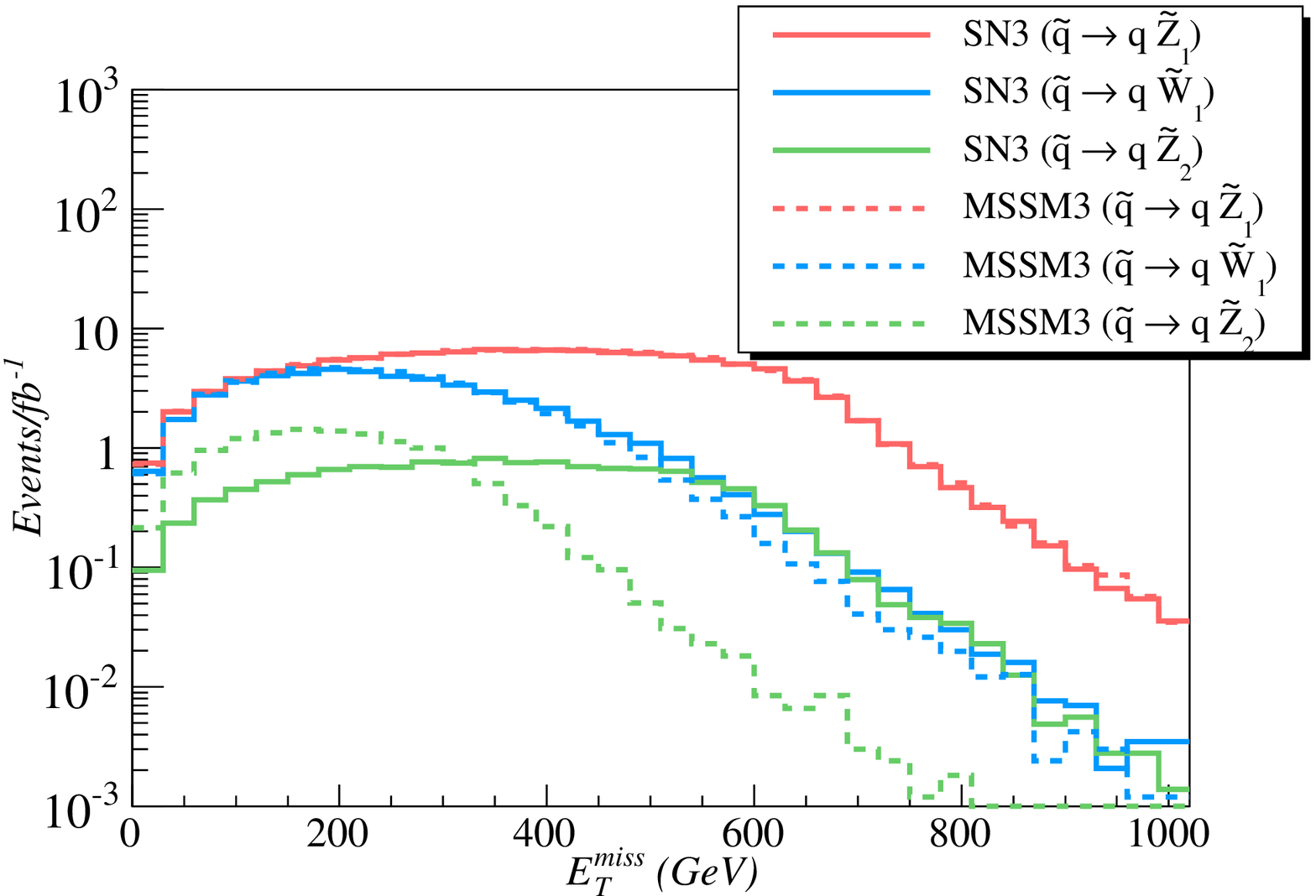}
\caption{Comparison of contributions to the $\eslt$ spectrum 
at $\sqrt{s}=7$~TeV for SN1 and MSSM1 (left) and for SN3 and MSSM3 (right);
SN2/MSSM2 gives almost the same picture as SN3/MSSM3.
}\label{fig:competmiss}}

Another general feature of the SNDM scenario is that, due to the invisible $\tz_2$ decay,
we expect a harder $\eslt$ spectrum as compared to a similar MSSM point.
This is illustrated in Fig.~\ref{fig:competmiss}, where we show the $\eslt$
distribution for the SN1/MSSM1 and SN3/MSSM3 points at detector level without any cuts (for details
on the event simulation, see next Section). Notice moreover 
that the subset of events with  $\tw_1\to l^\pm+{\rm LSP}$ has a slightly harder 
$\eslt$ spectrum in the SNDM case. This is because the light $\lsp$ LSP is more 
boosted than the $\tz_1$ LSP in the MSSM case; this difference 
becomes less significant if the LSP's mother already has a large boost. On the other
hand, the $\eslt$ spectrum coming from $\tq/\tg \to \tz_1$ events are identical in both models,
since the lightest neutralino decays invisibly in the SNDM.

Before studying the LHC phenomenology in more detail, 
a comment is in order concerning the assumption of flavor 
degeneracy.  
As discussed in \cite{Belanger:2010cd}, 
the exact mass splitting between the $\lsp$'s of different flavors strongly 
influences the DM-allowed parameter space. 
There is, however, no difference between one or three  
light sneutrinos if the mass splitting is $\agt 1$~GeV.
This is because a 1~GeV mass splitting 
is enough to suppress any co-annihilation contributions to $\Omega_{\tilde\nu}h^2$. 
Such a mass splitting is easily induced by small splittings in the 
soft terms, which are in fact rather generic even if one 
starts out with universal soft terms at a high scale. 
If we have, for instance, $m_{\tilde\nu_{1e,\mu}}>m_{\tilde\nu_{1\tau}}$, 
then the heavier $\tilde\nu_{1e,\mu}$ decays to the $\tilde\nu_{1\tau}$ LSP 
through 3-body modes~\cite{Kraml:2007sx}. The dominant decay is into neutrinos, while
visible decays (e.g., $\tilde\nu_{1e}\to e^\mp\tau^\pm\tilde\nu_{1\tau}$) 
have at most a few percent branching ratio. From the perspective of LHC  
signatures, the difference to the case of three exactly degenerate light 
sneutrinos is negligible. 
We therefore take 
$m_{\tilde\nu_{1e}}=m_{\tilde\nu_{1\mu}}=m_{\tilde\nu_{1\tau}}\equiv m_{\lsp}$ 
for simplicity. 
Note, however, that interesting non-trivial flavor structures beyond the scope of this study 
may appear in the presence of lepton-flavor violation~\cite{Kumar:2009sf,MarchRussell:2009aq}.

\section{Discovery Potential at LHC7}
\label{sec:Disc}

\renewcommand\tnu{\tilde\nu_1}

We now turn to the discovery potential of the SNDM model at the LHC, with
7 TeV CM energy and ${\cal O}(1)$ fb$^{-1}$ of integrated luminosity. 
The Monte Carlo
simulation details are presented in Sec.~\ref{sec:MC} and the main distinct
signatures are presented in Sec.~\ref{sec:lhc7dists}.

\subsection{Event Simulation} 
\label{sec:MC}

For the SM background, we include in our calculations all relevant $2 \to n$ processes
for the multi-lepton and multi-jet searches. Since in this Section we restrict
our results to the first LHC physics run ($\lesssim 1$~fb$^{-1}$ 
and $\sqrt{s} = 7$~TeV) we generate (at least) the equivalent
of 1~fb$^{-1}$ of events for each process, 
except for our QCD samples (see Table~\ref{table:bgs}).

For the simulation of the background events, we use {\tt AlpGen}~\cite{Alpgen}
to compute the hard scattering events and {\tt Pythia}~\cite{Pythia} for the
subsequent showering and hadronization.  For the final states containing
multiple jets (namely $Z(\to ll,\nu\nu) + jets$, $W(\to l\nu) + jets$,
$b\bar{b} + jets$, $t\bar{t} + jets$, $Z + b\bar{b} + jets$, $Z +
t\bar{t} + jets$, $W + b\bar{b} + jets$, $W + t\bar{t} + jets$ and QCD),
we use the MLM matching algorithm \cite{Alpgen} to avoid double counting.
All the processes included in our analysis are shown in
Table~\ref{table:bgs} as well as their total cross-sections and number of
events generated.  The SNDM spectrum
was generated with a modified version of {\tt Suspect}~\cite{Suspect}, 
which includes right-handed sneutrinos, while the decay table was computed
with {\tt CalcHEP}~\cite{Calchep}/{\tt micrOMEGAs2.4}\cite{Belanger:2006is,Belanger:2010cd}. 
Finally, using the SLHA~\cite{Skands:2003cj} 
interface, signal events were generated using {\tt Pythia6.4}~\cite{Pythia}.

For the event generation, we use a toy detector simulation with calorimeter cell size
$\Delta\eta\times\Delta\phi=0.05\times 0.05$ and rapidity $-5<\eta<5$. The HCAL
(hadronic calorimetry) energy resolution is taken to be
$80\%/\sqrt{E}+3\%$ for $|\eta|<2.6$ and FCAL (forward calorimetry) is
$100\%/\sqrt{E}+5\%$ for $|\eta|>2.6$, where the two terms are combined
in quadrature. The ECAL (electromagnetic calorimetry) energy resolution
is assumed to be $3\%/\sqrt{E}+0.5\%$. We use the {\tt Isajet}~\cite{isajet} cone-type 
 jet-finding algorithm  to group the hadronic
final states into jets. Jets and isolated leptons are defined as
follows: 
\bi
\item Jets are hadronic clusters with $|\eta| < 3.0$,
$R\equiv\sqrt{\Delta\eta^2+\Delta\phi^2}\leq0.4$ and $p_T(jet)>30$~GeV.
\item Electrons and muons are considered isolated if they have $|\eta| <
2.5$, $p_T(l)>10 $~GeV with visible activity within a cone of $\Delta
R<0.2$ about the lepton direction, $\Sigma p_T^{cells}<5$ GeV.  
\item  We identify hadronic clusters as 
$b$-jets if they contain a B hadron with $p_T(B)>15$~GeV, $\eta(B)<$ 3 and
$\Delta R(B,jet)< 0.5$. We assume a tagging efficiency of 60$\%$ and 
light quark and gluon jets can be mis-tagged
as a $b$-jet with a probability 1/150 for $p_T \leq 100$~GeV,
1/50 for $p_T \geq 250$~GeV, with a linear interpolation
for 100 GeV $\leq p_T \leq 250$~GeV. 
\ei

\begin{table}
\centering
\begin{tabular}{|l|c|c|}
\hline
                    &                 Cross   & number of \\
SM process & section & events \\
\hline
QCD: $2$, $3$ and $4$ jets (40 GeV$<p_T(j1)<100$ GeV) & $2.6\times 10^9$ fb  & 26M\\
QCD: $2$, $3$ and $4$ jets (100 GeV$<p_T(j1)<200$ GeV) & $3.9\times 10^8$ fb  & 44M\\
QCD: $2$, $3$ and $4$ jets (200 GeV$<p_T(j1)<500$ GeV) & $1.6\times 10^7$ fb  & 16M\\
QCD: $2$, $3$ and $4$ jets (500 GeV$<p_T(j1)<3000$ GeV) & $9.4\times 10^4$ fb  & 0.3M\\
$t\bar{t}$: $t\bar{t}$ + 0, 1 and 2 jets & $1.6\times 10^5$ fb&  5M\\
$b\bar{b}$: $b\bar{b}$ + 0, 1 and 2 jets & $8.8\times 10^7$  fb&  91M\\
$Z$ + jets: $Z/ \gamma (\to l\bar{l},\nu \bar{\nu})$ + 0, 1, 2 and 3 jets & $8.6\times 10^6$ fb&  13M\\
$W$ + jets: $W^{\pm} (\to l\nu)$ + 0, 1, 2 and 3 jets & $1.8\times 10^7$ fb&  19M\\
$Z$ + $t\bar{t}$: $Z/ \gamma (\to l\bar{l},\nu\bar{\nu})$ + $t\bar{t}$ + 0, 1 and 2 jets & $53$ fb &  0.6M\\
$Z$ + $b\bar{b}$: $Z/ \gamma (\to l\bar{l},\nu\bar{\nu})$ + $b\bar{b}$ + 0, 1 and 2 jets & $2.6\times 10^3$ fb  &  0.3M\\
$W$ + $b\bar{b}$: $W^{\pm} (\to all)$ + $b\bar{b}$ + 0, 1 and 2 jets & $6.4\times 10^3$ fb &  9M\\
$W$ + $t\bar{t}$: $W^{\pm} (\to all)$ + $t\bar{t}$ + 0, 1 and 2 jets & $1.8\times 10^2$ fb &  9M\\
$W$ + $tb$: $W^{\pm} (\to all)$ + $\bar{t}b(t\bar{b})$ & $6.8\times 10^2$ fb &  0.025M\\
$t\bar{t}t\bar{t}$ & $0.6$ fb &  1M\\
$t\bar{t}b\bar{b}$  & $1.0\times 10^2$ fb &  0.2M\\
$b\bar{b}b\bar{b}$ & $1.1\times 10^4$ fb &  0.07M\\
$WW$: $W^{\pm} (\to l\nu) + W^{\pm} (\to l\nu)$ & $3.0\times 10^3$ fb&  0.005M\\
$WZ$: $W^{\pm} (\to l\nu) + Z (\to all)$ & $3.4\times 10^3$ fb&  0.009M\\
$ZZ$: $Z (\to all) + Z (\to all)$ & $4.0\times 10^3$ fb&  0.02M\\
\hline
\end{tabular}
\caption{Background processes included in the discovery potential for LHC7, along with their 
total cross sections and number of generated events. All light (and {\it b}) partons in the 
final state are required to have $p_T> 40$~GeV. For QCD, we generate the hardest
final parton jet in distinct bins to get a better statistical representation of
hard events. For $Wtb$ production, additional
multi-jet production is only via the parton shower because the AlpGen 
calculation including all parton emission matrix elements
is not yet available. 
For this process, we apply the cut $|m(Wb)-m_t|\ge 5$~GeV
to avoid double counting events from real $t\bar{t}$ production.}
\label{table:bgs}
\end{table}

\subsection{Signal Distributions}
\label{sec:lhc7dists}

In Fig.~\ref{fig:sndmdists}a-c we show the $\eslt$, $n(l)$ and $n(j)$ distributions for the
benchmark points along with the SM background (BG) after the following set of cuts:
\bi
\item $\eslt > 400$ GeV, $n(j) > 3$, $p_T(j_1) > 150$ GeV, $p_T(j) > 50$ GeV and $S_T > 0.2$
\ei
where $S_T$ is the transverse sphericity and $p_T(j_1)$ is the $p_T$ of the hardest jet.

\FIGURE[t]{
\includegraphics[width=7cm]{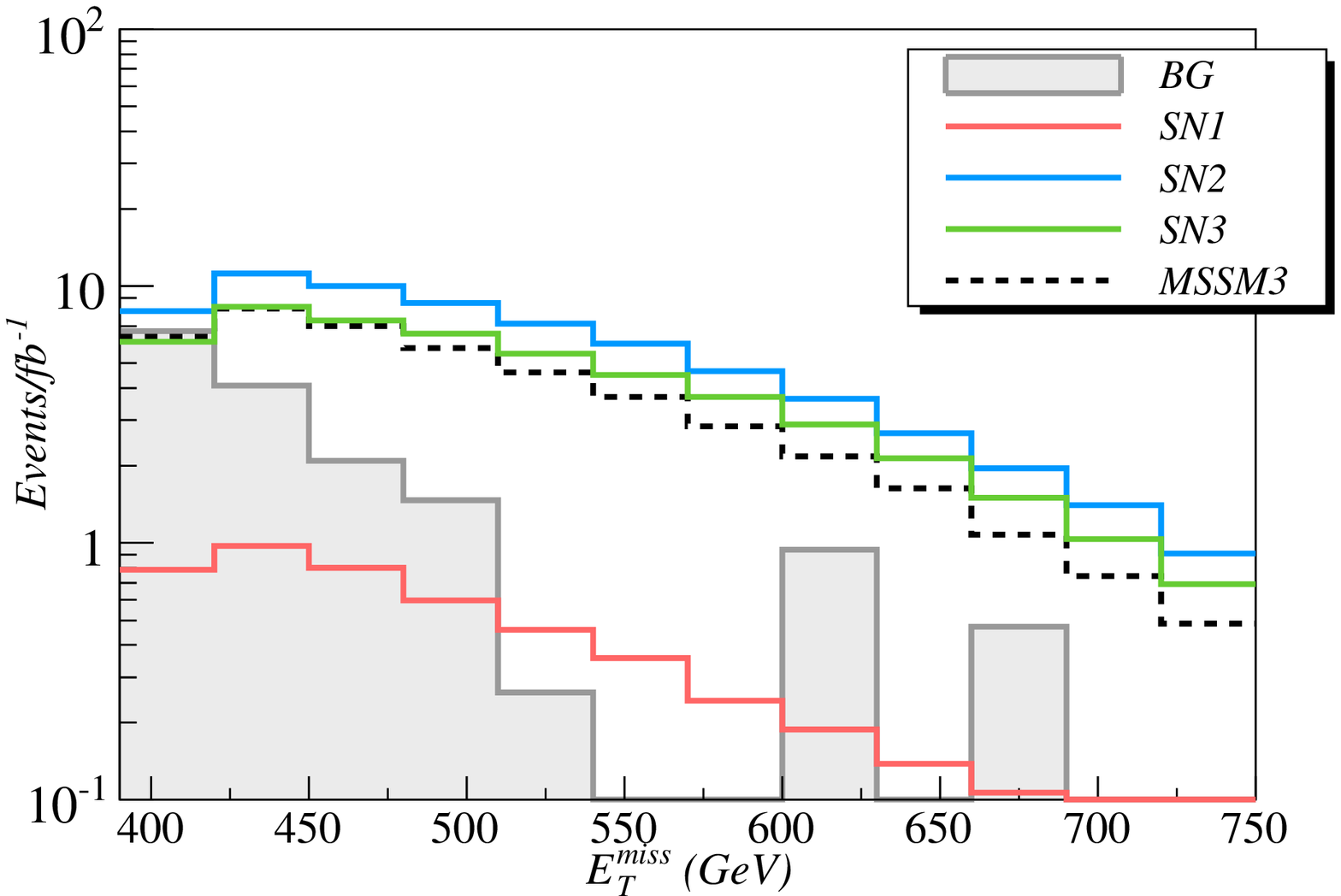}
\includegraphics[width=7cm]{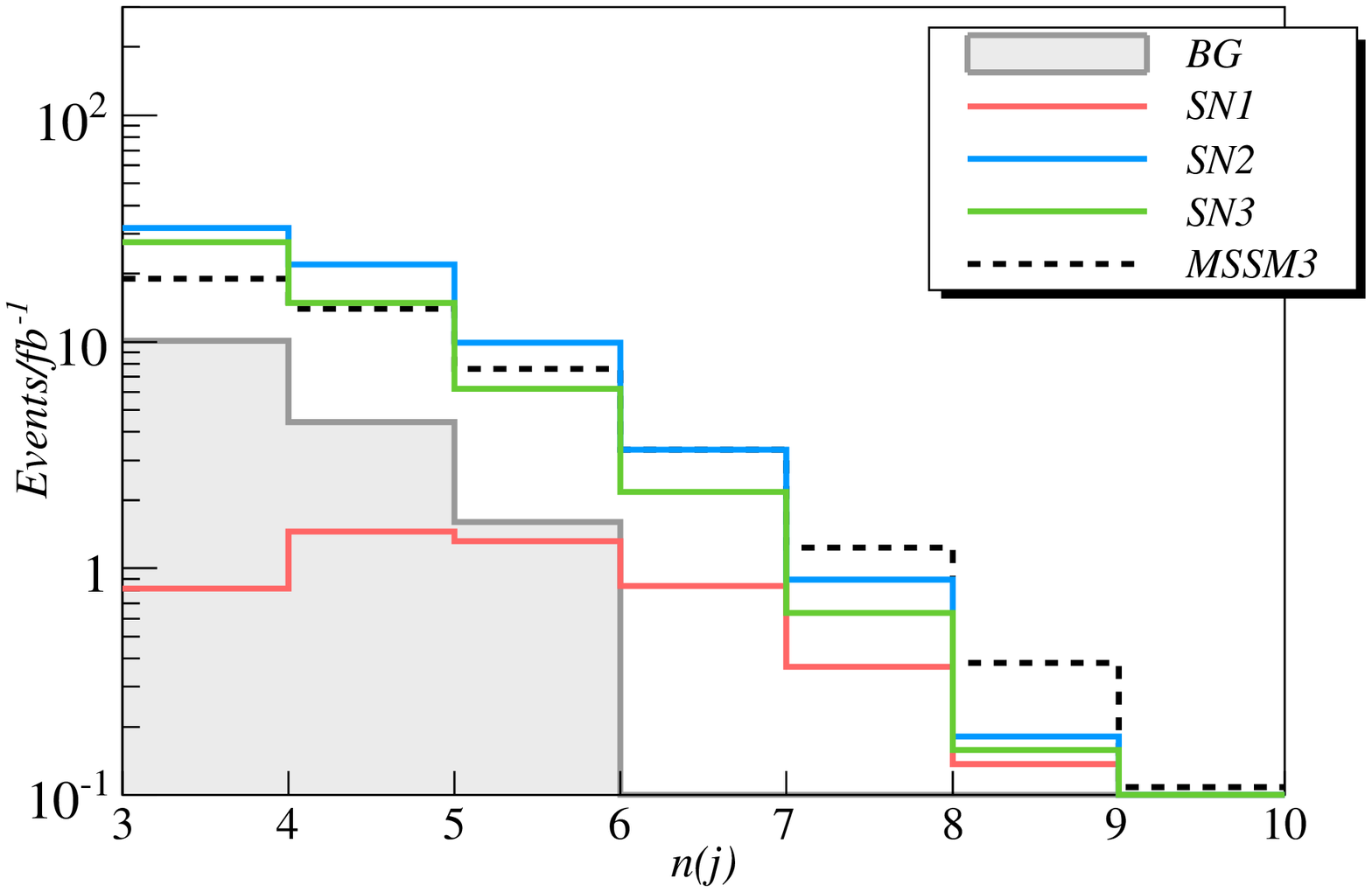}
\includegraphics[width=7cm]{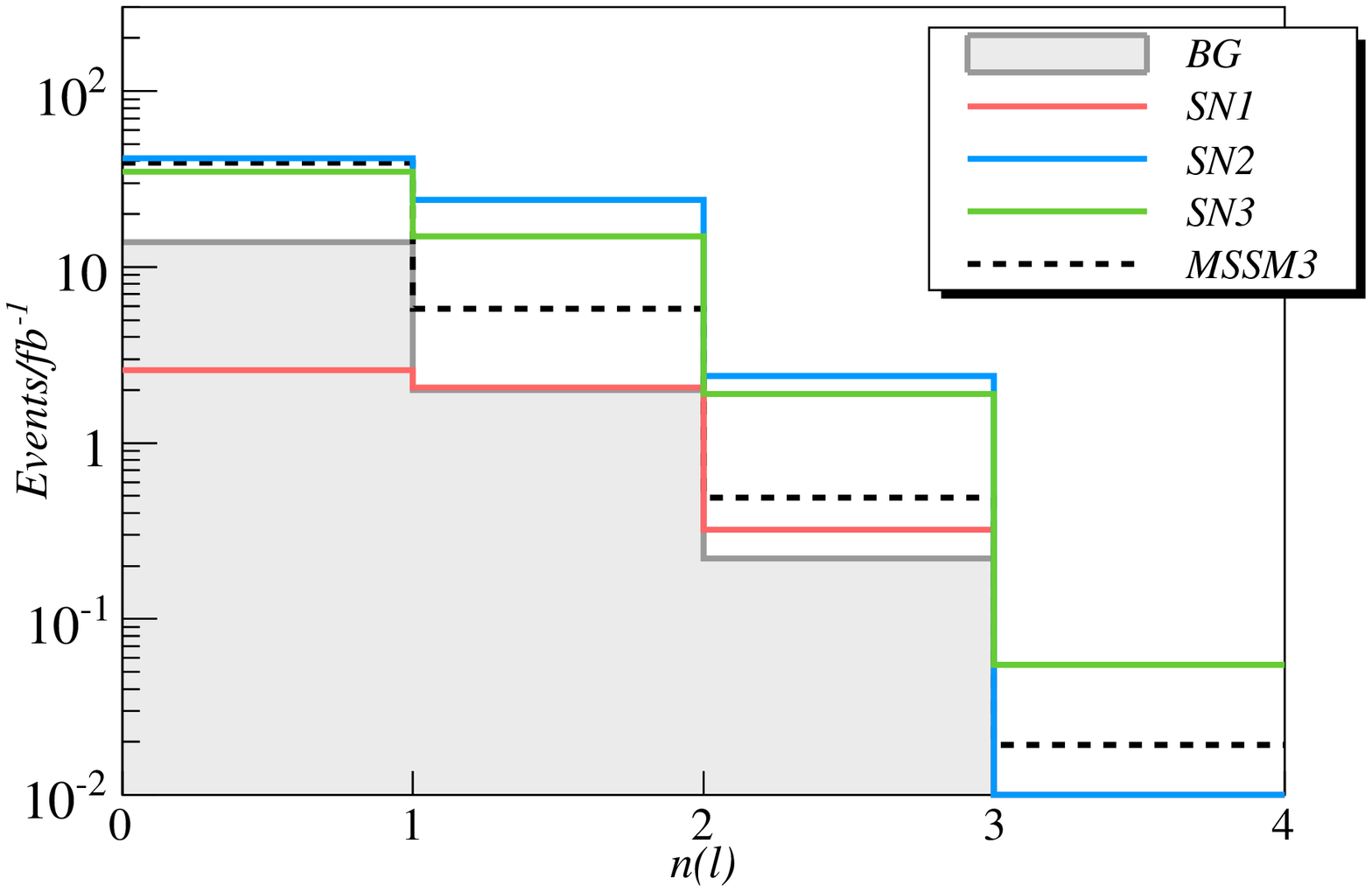}
\includegraphics[width=7cm]{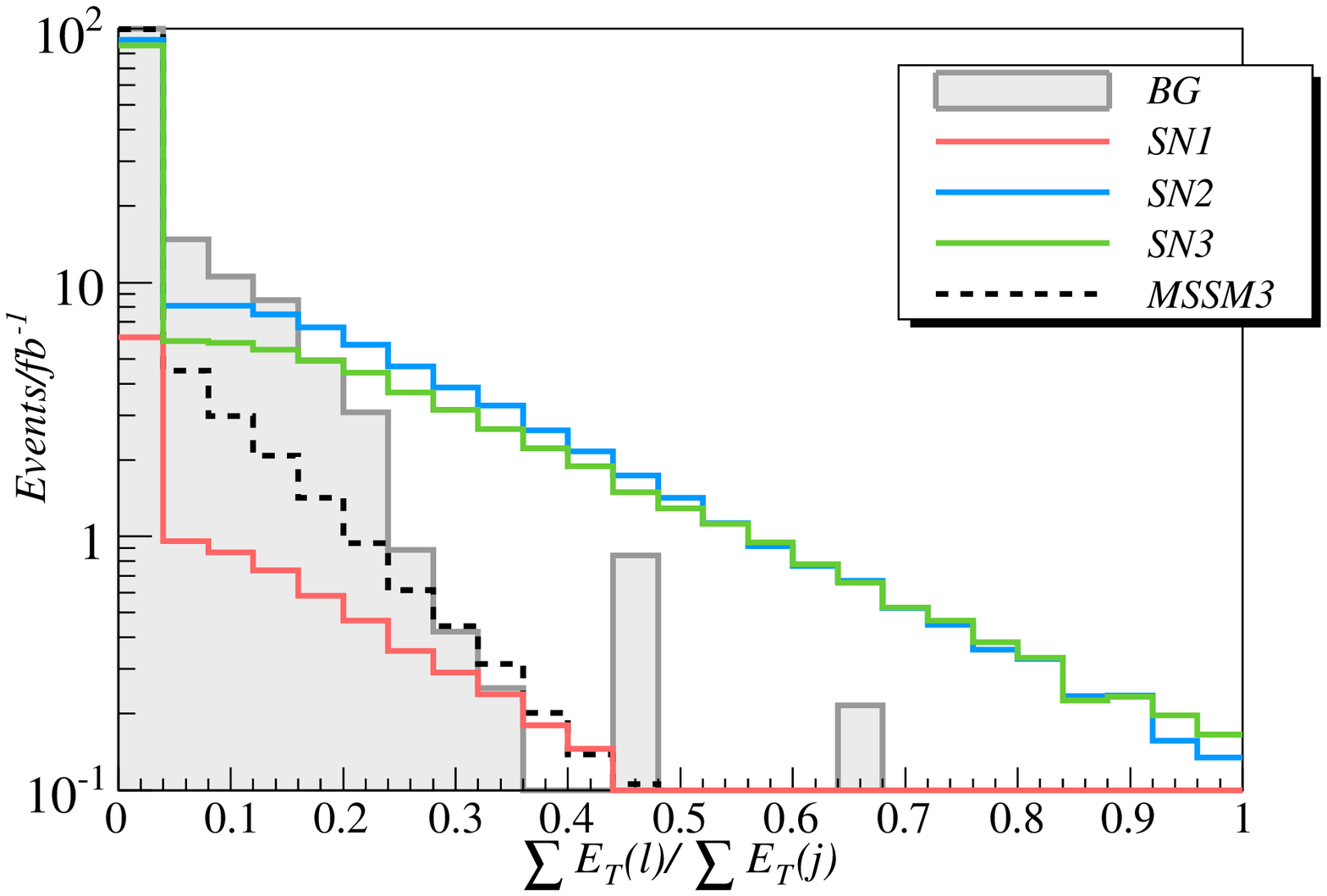}
\caption{ $\eslt$, $n(j)$, $n(l)$ and $\sum E_T(l)/\sum E_T(j)$
distributions for the model points in Table~\ref{tab:bm} along with SM background for LHC7. In frames
$a-c$ the following cuts have been
applied:  $\eslt > 400$ GeV, $n(j) > 3$, $p_T(j_1) > 150$ GeV, $p_T(j) > 50$ GeV and $S_T > 0.2$. While
frame $d$ has weaker cuts: $\eslt > 300$ GeV, $n(j) > 2$, $p_T(j_1) > 100$ GeV, $p_T(j) > 50$ and $S_T > 0.2$
}\label{fig:sndmdists}}

Despite having a softer $\eslt$ spectrum  than the SM BG,
the point SN1 has a much harder $n(j)$ distribution, which peaks at $n(j) = 4$, as expected
from the discussion in Sec.~\ref{sec:bench}.  As shown in Fig.~\ref{fig:sndmdists}b, the signal exceeds the SM BG in the 
$n(j) = 6, 7$ bins. Furthermore, with this set of cuts, the lepton number distribution for
point SN1 is already at the BG level for $n(l) = 1$, easily surpassing the BG
in the dilepton channel. However, due to its small cross section, the SN1 signal  requires several fb$^{-1}$
of integrated luminosity in order 
to become visible. We estimate that approximately 5(2)~fb$^{-1}$ are required
to claim a $5\sigma$($3\sigma$) evidence for the SN1 benchmark.

As mentioned in Sec.~\ref{sec:bench}, point SN2 has some distinct signatures from the
benchmark point SN1. While for the latter the sparticle production cross-sections is
dominated by $\tg\tg$, SN2 has the bulk of its signal coming from $\tq\tq$ and $\tq\tg$ events.
As a result the signal has a softer jet distribution, due to squark decays
to charginos and neutralinos, as shown by the decay branching ratios (BRs) 
in Table~\ref{tab:decays}. The same is valid for the SN3 signal, which is dominated by squark pair production.
This is clearly seen in the $n(j)$ distribution shown in Fig.~\ref{fig:sndmdists}b, once the overall signal
normalization is taken into account. On the other hand,
since the 2-body squark decay tends to produce boosted charginos and neutralinos, and because of the larger
production cross-section, 
points SN2 and SN3 have a harder $\eslt$ spectrum as compared to point SN1 (and the background), as shown in 
Fig.~\ref{fig:sndmdists}a.
Therefore, the LHC7 should be able to discover the SN2 and SN3 points
with hard $\eslt$ ($\gtrsim 300$~GeV) and $p_T(j_1)$ ($\gtrsim 250$~GeV) cuts in the $n(j) = 2$ or 3 channels. 
We estimate that a $5\sigma$ evidence for both SN2 and SN3 points can be achieved with 1~fb$^{-1}$ of integrated luminosity.
Furthermore, from Fig.~\ref{fig:sndmdists}c  we see that
both points are also visible in the dilepton channel.

Since the SNDM signal is rich in boosted leptons, we show in Fig.~\ref{fig:sndmdists}d the 
ratio between the scalar sum of the lepton and jet $E_T$'s, $E_T(l/j) \equiv \sum E_T(l)/ \sum E_T(j)$,
for a weaker set of cuts:
\bi
\item $\eslt > 300$ GeV, $n(j) > 2$, $p_T(j_1) > 100$ GeV, $p_T(j) > 50$ GeV and $S_T > 0.2$
\ei
As we can see, both the signal and BG distributions peak at low  $E_T(l/j)$ values.
However the SM BG falls much more sharply than the SN2 and SN3 signals, which are above background for
$E_T(l/j) \gtrsim 0.2$. Since the signal and SM distributions have very distinct shapes,
$E_T(l/j)$ can be used to discriminate between signal and background even for SNDM models
with smaller signal cross-sections. Nonetheless, point SN1 still is well below the SM BG to be seen at LHC7.

So far each of the SNDM signatures described above are common in standard MSSM scenarios,
including the CMSSM. Let us now discuss how the distributions shown in Fig.~\ref{fig:sndmdists} can help
to distinguish between the MSSM and the SNDM models. For this purpose we also show the respective
distributions for a MSSM case (MSSM3) with a spectrum identical to point SN3, but without 
RH sneutrino.\footnote{In order to avoid a proliferation of lines in the plots we do not show the distributions 
for MSSM1 and MSSM2, but we have checked that the general features discussed here by means 
of SN3 versus MSSM3 hold also for the other configurations.}  
As expected, MSSM3 has a softer  $\eslt$ spectrum than the SNDM point SN3, although the 
difference after cuts is not very large. 
Another distinction between the MSSM and SN points is their jet multiplicity.
In the SN case all charginos/neutralinos decay to $\tnu+ l/\nu$, while this channel
is absent for the MSSM points and neutralinos and charginos decay instead to the $\tz_1$ 
LSP plus $h$, $Z$ or $W$, see Sec.~\ref{sec:bench}. 
Therefore we expect SNDM models to have a softer $n(j)$ distribution when compared to a similar MSSM case. 
This is confirmed in Fig.~\ref{fig:sndmdists}b, which shows that the $n(j)$ distribution for the MSSM3 model
is suppressed in the $n(j)=3$ bin and enhanced in the higher bins when compared
to the SN3 point (once the total signal normalization is taken into account).

The main distinct feature of the SNDM scenario appears, nevertheless, in the multilepton distribution. As discussed in Sec.~\ref{sec:bench},
SNDM models are expected to be rich in hard leptons coming almost exclusively from $\tw_1$ decays. 
As a result, the leptons in dilepton events will have uncorrelated flavors and kinematics. Furthermore, higher lepton multiplicities
only appear, at much smaller rates, from top decays. In corresponding MSSM scenarios, 
$\tw_1$'s and $\tz_2$'s decays to leptons are sub-dominant: in case of $\tw_1$, they 
are given by the BRs of the $W$, while in case of $\tz_2$, they are suppressed by the 
high BR into $h^0$. 
To give concrete numbers, the ratio of dilepton/0-lepton events is $\sim 0.12,\ 0.06$ and $0.06$
for points SN1, SN2 and SN3, respectively, see Fig.~\ref{fig:sndmdists}c, 
while for the MSSM1, MSSM2 and MSSM3 points it is 
 $\sim 0.04,\ 0.01$ and $0.01$.
Furthermore, dilepton events in MSSM models often come from $\tz_2$ decays to $l^+l^-\tz_1$ and 
hence consist of same-flavor OS dileptons.
Therefore the ratio of OS and SS dilepton events is another useful discriminant.
For the set of cuts used in Fig.~\ref{fig:sndmdists}c we have SS/OS = $0.94$, $0.94$ and $0.62$
for points SN1, SN2 and SN3, respectively. The SS/OS ratio is considerably smaller for SN3 
than for the other points, because its signal is dominated by squark pair production, leading to 
less SS dileptons. Although the SN2 point
also has a large $\tq \tq$ production cross-section,
the $n(j) \geq 3$ cut suppresses the contribution from $\tq \tq$ events to the SN2 signal, enhancing the SS/OS ratio in this case.
For the corresponding MSSM models and the same set of cuts, we have SS/OS = $0.75$, $0.54$ and $0.39$ for MSSM1, MSSM2 and MSSM3, respectively.
As we can see, the SS/OS ratio is significantly enhanced in SNDM models, when compared to the MSSM.
Furthermore, as seen in Fig.\ref{fig:sndmdists}d, the $E_T(l/j)$ 
distribution for the MSSM is much softer than in the SNDM case, which, together with the $n(j)$ and $n(l)$ distributions
can also point to a light sneutrino LSP.

We conclude that, while the $\eslt$ and $n(j)$ distributions are more model dependent and more sensitive to NLO and detector systematics, the
$n(l)$ and $E_T(l/j)$ distributions are promising candidates for an early distinction between the MSSM and SNDM cases.
Although the first LHC run will probably not accumulate enough luminosity to make use of the dilepton invariant mass, depending on
the signal cross-section, the dilepton  channel may already indicate which is the relevant model.
However, a more decisive answer will have to wait for the later physics run with higher energy. In the next section we discuss how LHC14 ($\sqrt{s}=14$~TeV) 
with an integrated luminosity of 100 fb$^{-1}$ can give information on the SNDM spectrum and thefore provide stronger evidence that the SNDM case is realized.

\section{Measuring Masses at LHC14}
\label{sec:Masses}

Despite the distinct features of the SNDM model discussed in the previous sections,
information on the sparticle spectrum can give a more decisive evidence for
sneutrino DM. While in most MSSM models,
a neutralino LSP consistent with DM and collider constraints implies $m_{\tz_1} \gtrsim 50$ GeV
(assuming gaugino mass unification), the SNDM model has a much lighter LSP.
Since several fb$^{-1}$ are required for any mass measurement, from now on we will present
all our results for LHC14 ($\sqrt{s} = 14$ TeV and $\mathcal{L} = 100$ fb$^{-1}$).
We include both SUSY and SM backgrounds, which include all the relevant processes
listed in Table~\ref{table:bgs}, but at $\sqrt{s} = 14$ TeV (for more details
see \cite{Baer14}).

\subsection{Method}
Recently much improvement has been made on mass measurement methods at hadron colliders and several
distinct mass measurement techniques have been suggested, such as
kinematic endpoints and shapes (invariant mass, $s_{min}$, ...)
and transverse mass methods  ($m_{T2}$, $m_{CT}$,...).
See \cite{Matchev:2009iw,Barr:2011xt} for comprehensive reviews and references.
Here we adopt
the subsystem $m_{T2}$ method, described in detail in Ref.~\cite{Mt2sub}.
It consists in applying the $m_{T2}$ endpoint technique to particular 
subsets of the event topology. 

\subsubsection*{Gluino-pair production}

First, we consider gluino-pair production, with each gluino going 
through the cascade decay:
\begin{equation}
  \tg \to qq + \tw_1 \to qq + l + \tnu \label{eq:evtop1}
\end{equation}
From this event topology we can form three subsystems:
\begin{equation}
  \begin{array}{llcl}
      & \bf P^{(1)}+P^{(2)} & \to & \bf vis^{(1)}+ D^{(1)}+ vis^{(2)}+D^{(2)} \\[1mm]
   {\rm 1.}\; & \tg^{(1)}+\tg^{(2)} & \to & qql^{(1)}+\tnu^{(1)}+qql^{(2)}+\tnu^{(2)} \\ 
   {\rm 2.} & \tg^{(1)}+\tg^{(2)} & \to & qq^{(1)} +l\tnu^{(1)}+qq^{(2)}+l\tnu^{(2)}\\
   {\rm 3.} & \tw_1^{(1)}+\tw_1^{(2)} &\to& l^{(1)}+\tnu^{(1)}+l^{(2)}+\tnu^{(2)} 
  \end{array}\label{eq:subs}
\end{equation}
where the upper index labels the decay branch and the final states are grouped into a visible (vis) and a daughter (D) component.
For instance, in the first case the gluinos are the parents (P), while $qql$ and $\tnu$ are the visible and daughter components,
respectively. On the other hand, for the third subsytem, the charginos are the parents and $l$ and $\tnu$ are the visible and daughter
components. The (subsystem) $m_{T2}$ variables are then constructed as defined in \cite{Lester1999, Barr2003}, but using the visible and daughter
components defined within each subsystem:
\begin{equation}
  m_{T2}(m_x) = \min_{\pT({\rm D}_1) + \pT({\rm D}_2) = - \pT({\rm vis}_1) - \pT({\rm vis}_2)} [{\rm max}(m_T^{(1)},m_T^{(2)})]
\end{equation}
where
\begin{equation} 
  m_T^{(i)} = \sqrt{m_{{\rm vis}_i}^2 + m_x^2 + 2(E_T({\rm vis}_i) E_T({\rm D}_i) - \pT({\rm vis}_i).\pT({\rm D}_i))} \end{equation}
and $m_x$ is the trial daughter mass. Since $m_{T2}(m_x = m_{\rm D}) \leq m_{\rm P}$, the 
value of $m_{T2}^{max}(m_{\rm D})$ determines the parent's mass in each subsystem. For the above example we have:
\begin{enumerate}
	\item $m_{T2}^{qql,max}(m_x = m_{\tnu}) = m_{\tg}$
	\item $m_{T2}^{qq,max}(m_x = m_{\tw_1}) = m_{\tg}$
	\item $m_{T2}^{l,max}(m_x = m_{\tnu}) = m_{\tw_1}$
\end{enumerate}
where $m_{T2}^{qql}$, $m_{T2}^{qq}$ and $m_{T2}^{l}$ are the $m_{T2}$ subsystem variables
for the subsystems 1, 2 and 3 defined in Eq.~(\ref{eq:subs}), respectively. Note that for the second
subsystem the daughter is defined as $l\tnu$, which has invariant mass $m_{\tw_1}$.  
However, as the above relations show, in order to obtain the parent's mass, the daughter
mass has to be known. The strength of the $m_{T2}$ method comes from the fact that
analytical expressions are known for the function $m_{T2}^{max}(m_x)$ and
can be used to simultaneously extract both the parent and daughter masses from data. For the subsystems 1 and 2
in Eq.~(\ref{eq:subs}), we have:
\begin{equation}
m_{T2}^{qql,max}(m_x) = \left\{ 
\begin{array}{lr}
 \frac{m_{\tg}^2 - m_{\tnu}^2}{2 m_{\tg}} + \sqrt{\left(\frac{m_{\tg}^2 - m_{\tnu}^2}{2 m_{\tg}}\right)^2 + m_x^2} &\mbox{, if $m_x < m_{\tnu}$}\\
m_{\tg}\left(1-\frac{m_{\tw_1}^2}{2m_{\tg}^2} -\frac{m_{\tnu}^2}{2m_{\tw_1}^2}\right) + \sqrt{m_{\tg}^2\left(\frac{m_{\tw_1}^2}{2m_{\tg}^2}-\frac{m_{\tnu}^2}{2m_{\tw_1}^2}\right)^2 + m_x^2} &\mbox{, if $m_x > m_{\tnu}$}
\end{array} \right. \label{eq:mt2qql}
\end{equation}
and 
\begin{equation} 
m_{T2}^{qq,max}(m_x) = \left\{ 
\begin{array}{lr}
 \frac{m_{\tg}^2 - m_{\tw_1}^2}{2 m_{\tg}} + \sqrt{\left(\frac{m_{\tg}^2 - m_{\tw_1}^2}{2 m_{\tg}}\right)^2 + m_x^2} &\mbox{, if $m_x < m_{\tw_1}$}\\
  m_{\tg} - m_{\tw_1} + m_x &\mbox{, if $m_x > m_{\tw_1}$}
\end{array} \right. \label{eq:mt2qq} 
\end{equation}

The above expressions were derived in Refs.~\cite{Cho2007,Mt2sub} under the assumption of no initial state radiation (ISR), so the total parent $p_T$ ($|\pT(\tg^{(1)})+\pT(\tg^{(2)})|$) is zero. In Ref.~\cite{Mt2sub} it is shown that, unless $p_{T} \gtrsim m_{\tg}$, this is a reasonable approximation. Therefore for $m_{\tg} \gtrsim 700$ GeV, we can safely neglect the ISR effect. On the other hand, for the subsystem 3, the parent system ($\tw_1^{(1)}+\tw_1^{(2)}$) is expected to have large $p_T$, since they are produced from gluino decays and are much lighter than their parents ($m_{\tw_1} = 227$ GeV for the cases we consider). In this case we have to include the transverse momentum effect in 
$m_{T2}^{l,max}(m_x)$\cite{Mt2sub}:

\begin{equation}
m_{T2}^{l,max}(m_x) = \left\{ 
\begin{array}{lr}
 \sqrt{\left(\mu_- + \sqrt{(\mu_- + \frac{p_T}{2})^2 + m_x^2}\right)^2 -\frac{p_T^2}{4}} &\mbox{, if $m_x < m_{\tnu}$}\\
   \sqrt{\left(\mu_+ + \sqrt{(\mu_+ - \frac{p_T}{2})^2 + m_x^2}\right)^2 -\frac{p_T^2}{4}} &\mbox{, if $m_x > m_{\tnu}$}
\end{array} \right. \label{eq:mt2l}
\end{equation}
where $\mu_{\pm} = \frac{m_{\tw_1}^2 - m_{\tnu}^2}{2 m_{\tw_1}} \left(\sqrt{1 + \frac{p_T^2}{4 m_{\tw_1}^2}} \pm \frac{p_T}{2 m_{\tw_1}}\right)$
and $p_T$ is the chargino pair total transverse momentum. Note that, neglecting initial state radiation for the gluino pair, we have
\begin{equation}
p_T = |\vec{p}_{T}(q_1)+\vec{p}_{T}(q_2)+\vec{p}_{T}(q_3)+\vec{p}_{T}(q_4)| \equiv p_T^{trans}
\end{equation}
Therefore, selecting events with a fixed $p_T^{trans}$,
 Eq.~(\ref{eq:mt2l}) can be used to obtain $m_{\tw_1}$ and $m_{\tnu}$. However this cut considerably
reduces the statistical significance of the $m_{T2}$ distribution.

\subsubsection*{Squark production}

The decay chain (\ref{eq:evtop1}) is relevant
for point SN1, where the signal is dominated by gluino pair-production, while for points SN2 and SN3,
the sparticle production cross-section is dominated by squark pair production or squark-gluino production.
In this case we will consider the following event topology:
\begin{equation}
\tq_L \to q + \tw_1 \to q + l + \tnu \label{eq:evtop2}
\end{equation}
The subsystems in this case are analagous to the
ones defined in Eq.~(\ref{eq:subs}), but with one less quark in the final state.
Using the same notation as before, we label them as
$m_{T2}^{ql}$, $m_{T2}^{q}$ and $m_{T2}^{l}$, with:
\begin{enumerate}
	\item $m_{T2}^{ql,max}(m_x = m_{\tnu}) = m_{\tq}$
	\item $m_{T2}^{q,max}(m_x = m_{\tw_1}) = m_{\tq}$
	\item $m_{T2}^{l,max}(m_x = m_{\tnu}) = m_{\tw_1}$
\end{enumerate}
The new $m_{T2}^{max}$ functions now obey~\cite{Cho2007,Mt2sub}:
\begin{equation}
m_{T2}^{ql,max}(m_x) = \left\{ 
\begin{array}{lr}
 \frac{m_{\tq}^2 - m_{\tnu}^2}{2 m_{\tq}} + \sqrt{\left(\frac{m_{\tq}^2 - m_{\tnu}^2}{2 m_{\tq}}\right)^2 + m_x^2} &\mbox{, if $m_x < m_{\tnu}$}\\
m_{\tq}\left(1-\frac{m_{\tw_1}^2}{2m_{\tq}^2} -\frac{m_{\tnu}^2}{2m_{\tw_1}^2}\right) + \sqrt{m_{\tq}^2\left(\frac{m_{\tw_1}^2}{2m_{\tq}^2}-\frac{m_{\tnu}^2}{2m_{\tw_1}^2}\right)^2 + m_x^2} &\mbox{, if $m_x > m_{\tnu}$}
\end{array} \right. \label{eq:mt2ql}
\end{equation}

\begin{equation} 
m_{T2}^{q,max}(m_x) = \left\{ 
\begin{array}{lr}
 \frac{m_{\tq}^2 - m_{\tw_1}^2}{2 m_{\tq}} + \sqrt{\left(\frac{m_{\tq}^2 - m_{\tw_1}^2}{2 m_{\tq}}\right)^2 + m_x^2} 
\end{array} \right. \label{eq:mt2q}
\end{equation}
while  $m_{T2}^{l,max}(m_x)$ is identical to Eq.~(\ref{eq:mt2l}). Note that,
for a pair of $m_x$ values ($m_x = 0$ and $m_x \gg m_D$), the gluino subsystems 
provide 6 constraints on the 3 masses involved, while the squark subsystems provide 5 constraints
on 3 masses\footnote{If, as in Eq.~(\ref{eq:mt2l}), we assume
a non-zero transverse momentum for the squark pair, Eq.~(\ref{eq:mt2q}) would have two branches and could provide
an extra constraint on the masses. However, as in the gluino case, ISR effects for squark pair production
are negligible.}. Therefore in both cases there is an arbitrariness on how to extract the mass values.
Since the $m_{T2}^{l}$ distribution requires an extra cut on $p_T$ in order to allow for the determination
of $m_{\tw_1}$ and $m_{\tnu}$, it will necessarily have a much smaller statistical significance than the other
$m_{T2}$ distributions. Hence we will make use only of the first two subsystems, which
are already sufficient to determine all the masses involved in the process. For that, we adopt the following procedure:
\bi
\item For discrete values of $m_x$, we extract the value of $m_{T2}^{max}(m_x)$ for the subsystems 1 and 2,
from the respective $m_{T2}$ distributions, following the algorithm defined in Appendix \ref{sec:extr};
\item We then simultaneously fit the results to the appropriate $m_{T2}^{max}$ functions,  Eqs.~(\ref{eq:mt2qql}), (\ref{eq:mt2qq}), (\ref{eq:mt2ql}) or (\ref{eq:mt2q});
\item The mass values and their uncertainties are then extracted from the best fit result.
\ei

Although only two $m_{T2}^{max}$ measurements at two different $m_x$ values are already sufficient
to obtain all masses, fitting the $m_{T2}^{max}(m_x)$ expressions for a wide range of $m_x$ values has two main advantages. 
First, it is less sensitive to uncertainties in extracting $m_{T2}^{max}$ from the $m_{T2}$ distributions. Second,
it allows us to test our underlying model asssumptions since a poor fit would indicate that the events selected
do not correspond to the topologies (\ref{eq:evtop1}) or (\ref{eq:evtop2}).

\subsubsection*{Zero-lepton channel}

In principle we can also determine the $\tz_1$ and $\tz_2$ masses
if we look at the 0 lepton plus $\eslt$ channel.
For point SN1 this channel is dominated by gluino decays to neutralinos:
\begin{equation}
\tg \to qq + \tz_1/\tz_2 \to qq + \nu + \tnu
\end{equation}
The $m_{T2}^{max}(m_x)$ function for this process obeys Eq.~(\ref{eq:mt2qq}) with $m_{\tw_1} \to m_{\tz_1,2}$. In this case,
the $m_{T2}$ distribution (for a fixed $m_x$) would present two endpoints, one from $\tg \to qq + \tz_2$ and one
from $\tg \to qq + \tz_1$. Since the latter will necessarily be at higher $m_{T2}$ values, a simple extraction
of $m_{T2}^{max}$ will always give the $\tz_1$ endpoint, with the $\tz_2$ endpoint partially obscured by
the $m_{T2}$ distribution from $\tz_1$ events. We have verified that, for the SN1 point, the signal/BG
ratio is too small in the 4 jets, 0 lepton plus $\eslt$ channel, 
making such a measurement unviable for $\mathcal{L} \sim 100$~fb$^{-1}$.

For points SN2 and SN3 the 0 lepton plus $\eslt$ channel is dominated by $\tq$ decays:
\begin{equation}
\tq \to q + \tz_1/\tz_2 \to q + \nu + \tnu
\end{equation}
and the $m_{T2}^{max}(m_x)$ function obeys Eq.~(\ref{eq:mt2q}) with $m_{\tw_1} \to m_{\tz_1,2}$.
Since BR$(\tq_R \to \tz_1) \simeq 100\%$, while BR$(\tq_L \to \tz_2) \simeq 33\%$, the 2 jets, 0 lepton plus $\eslt$
channel comes mainly from $\tq_R$ decays and will have a large cross section.
Therefore in this case it is possible to extract $m_{\tz_1}$, once $m_{\tq_R}$ is known. Hence, after determining
the $m_{\tq}$, $m_{\tw_1}$ and $m_{\tnu}$ values from $\tq \to \tw_1 + q$ decays, we will use the zero lepton channel
to determine $m_{\tz_1}$ under the assumption that left and right-handed squarks are degenerate.

The results obtained applying the procedure described here to the benchmark points SN1--SN3 are 
presented below in sections~\ref{mt2results1} and \ref{mt2results2}. 
In our analysis we include all relevant SUSY and SM backgrounds,
as well as detector and combinatorics systematics. We always assume $\sqrt{s} = 14$ TeV and $\mathcal{L} = 100$ fb$^{-1}$.

\subsection{Results for SN1}
\label{mt2results1}

\FIGURE[t]{
\includegraphics[width=10cm]{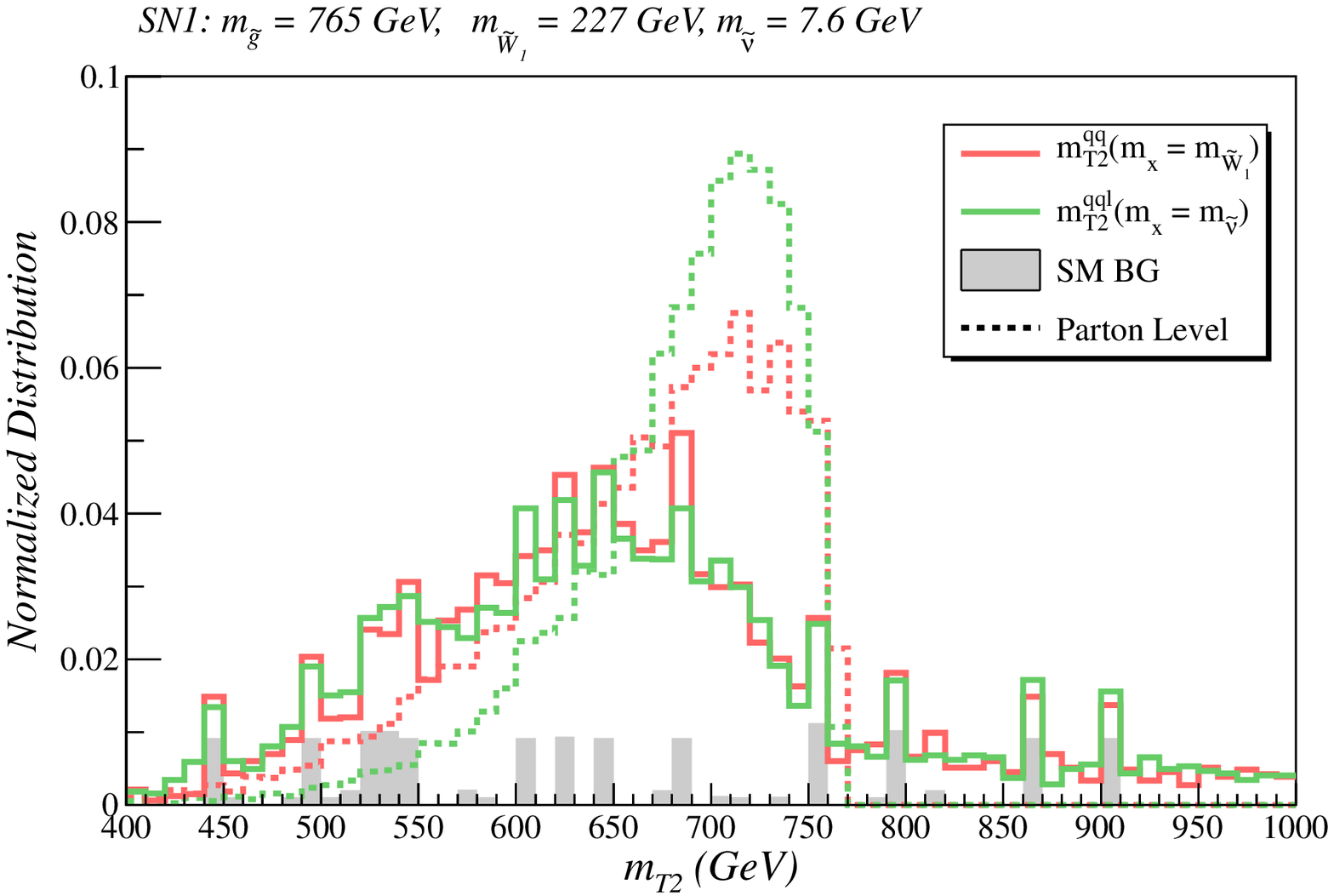}
\caption{The $m_{T2}$ distributions for the first two
subsystems defined in Eq.~(\ref{eq:subs}) for point SN1. The dashed lines
show the signal distributions at parton level, the shaded histogram shows the SM contribution and
the solid lines show the signal plus background distributions at the detector level,
after the cuts Eq.~(\ref{eq:cuts1}) have been applied. 
All distributions are normalized to unity.
}\label{fig:mt2}}

To apply the $m_{T2}$ subsystem method for the SN1 point, we consider the 
event topology Eq.~(\ref{eq:evtop1}),
which can be selected with the following set of cuts:
\begin{equation}
  \begin{array}{l}
     \eslt > 100\;{\rm GeV},~ n(j) = 4,~ n(l) = 2,~ n(b) = 0,\\ 
     p_T(j_1) > 100\;{\rm GeV},~ p_T(j) > 50\;{\rm GeV},~ p_T(l) > 30\;{\rm GeV}\,. 
  \end{array} \label{eq:cuts1}
\end{equation}
Furthermore, to reduce the $Z/\gamma + jets$ background, we veto OS-dilepton events with 
\be
80\;{\rm GeV} < m(l^+l^-) < 100\;{\rm GeV\;\; or }\;\;m(l^+l^-) < 40\;{\rm GeV}\,.
\ee
The total SM BG for this set of cuts is 0.7 fb, while the SN1 signal is 4.6 fb. Due to
the small BG level, our MC samples have only a few events, dominated by $t\bar{t} + 2 jets$.
Although we include the full SM background in our results, the effect is subdominant.
A more relevant issue is the jets and
lepton combinatorics. For the subsystem 1 (2), defined in Eq.~(\ref{eq:subs}), it is necessary to group the 4 jets and 2 leptons (4 jets)
into 2 visible groups, $vis^{(1)}$ and $vis^{(2)}$.
Several methods have been proposed to deal with the combinatorics issue~\cite{Cho2007,Rajaraman2010},
which usually rely on kinematical correlations between the final states. However, since we are only interested
in obtaining $m_{T2}^{max}$, for each event we select the grouping which gives the minimum
$m_{T2}$ value for that event. This way the $m_{T2} < m_{T2}^{max}(m_x)$ relation is still preserved even if the wrong grouping
is selected. Nevertheless, initial and final state radiation (FSR), signal background and detector energy smearing
 still affects the $m_{T2}$ distribution, resulting in a tail for values above $m_{T2}^{max}(m_x)$.

In Fig.~\ref{fig:mt2}, we present the $m_{T2}^{qql}$ and $m_{T2}^{qq}$ distributions for the SN1 signal plus the SM background,
where the trial mass was chosen as the respective daughter mass. For comparison purposes we also present the exact parton
level distributions (dashed lines) for the signal. The spikes in the (solid) $m_{T2}$ distributions come from 
MC fluctuations in the SM background. As can be seen from Fig.~\ref{fig:mt2}, both distributions have an edge at
\bi
      \item $m_{T2}^{qql} \sim 760$ GeV and $m_{T2}^{qq} \sim 760$ GeV\,.
\ei
The above values agree well with the expected value, $m_{\tg}$.
Figure~\ref{fig:mt2} also shows that the $m_{T2}^{qql}$ and $m_{T2}^{qq}$ distributions are strongly
affected by the cuts, ISR, FSR and energy smearing effects, when compared to the parton level distributions.
Furthermore, our solution
to the combinatorics usually shifts the $m_{T2}$ distribution to lower values,
what diminishes the peak, resulting in a less evident edge.

\FIGURE[t]{
\includegraphics[width=7.3cm]{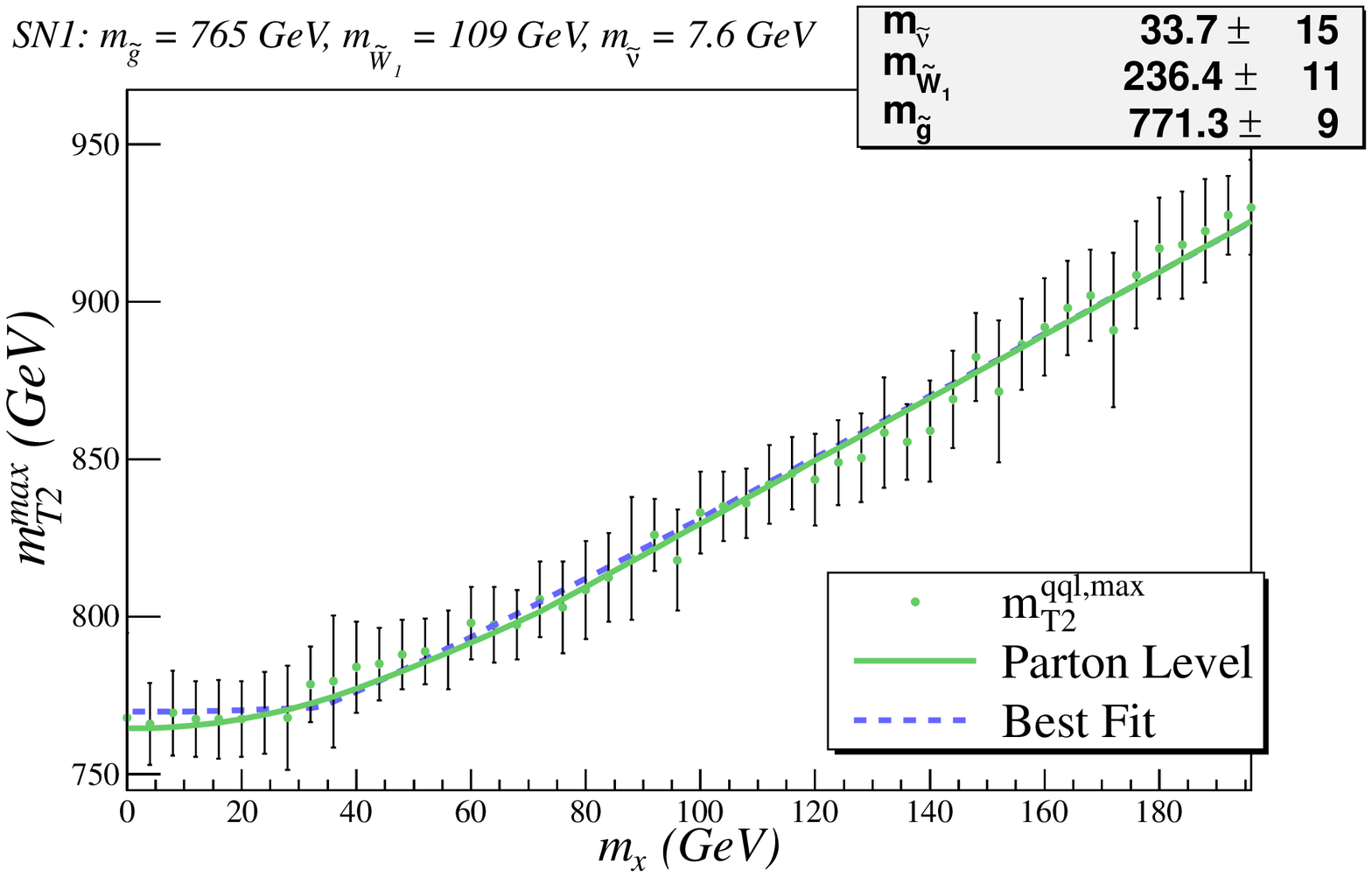}
\includegraphics[width=7.3cm]{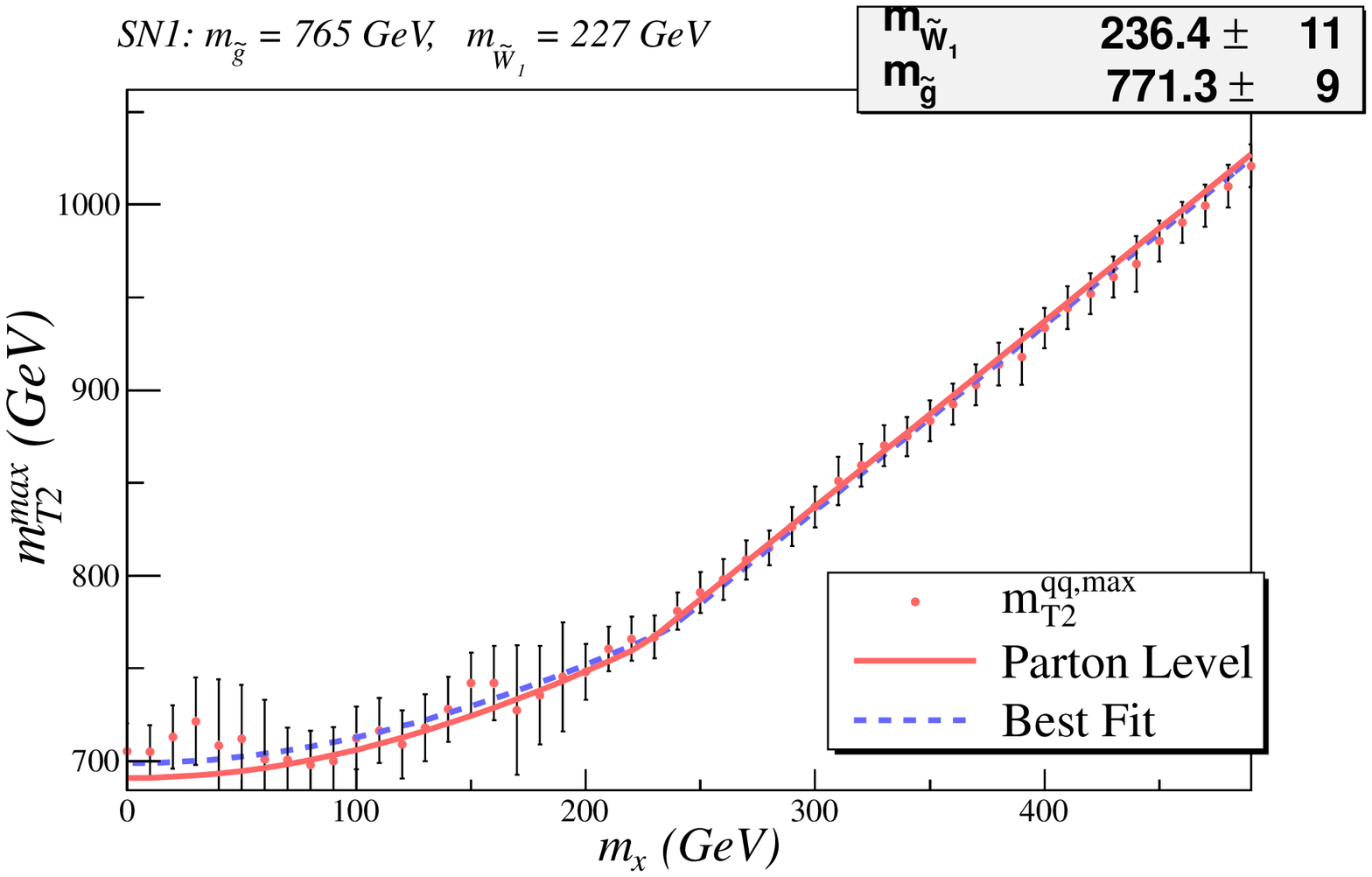}
\caption{$m_{T2}^{max}$ for point SN1, as a function of the trial daughter mass $m_x$,
for the first two subsystems defined in Eq.~(\ref{eq:subs}). The dashed blue lines show the best fit result obtained
using Eqs.~(\ref{eq:mt2qql})--(\ref{eq:mt2qq}). The solid lines show the exact result obtained at
parton level.
}\label{fig:mt2max}}

The distributions in Fig.~\ref{fig:mt2} assume a specific value for the trial mass $m_x$, which was chosen to
be the exact daughter mass in each case. However, as discussed in Sec.\ref{sec:Masses}, the value of
$m_{T2}^{max}$ as a function of $m_x$ allows us to extract both the parent and daughter masses, through
Eqs.~(\ref{eq:mt2qql}) and (\ref{eq:mt2qq}). Figure~\ref{fig:mt2max} shows the results obtained from  fitting the $m_{T2}^{max}(m_x)$ functions
to the $m_{T2}^{max}$ values extracted from the simulated data using the algorithm outlined in the Appendix.
As can be seen from Figs.~\ref{fig:mt2max}a and b,
the best fit for the $m_{T2}^{qql,max}$ and $m_{T2}^{qq,max}$ curves (dashed blue lines) 
agree well with the exact solution (solid lines). 
The final result for the three masses, taken as the best simultaneous fit to both $m_{T2}^{max}$ curves, is shown
in Table~\ref{table:masses}. The approximate precision for $m_{\tg}$ and $m_{\tw_1}$ is under a few percent,
while the precision for the $\lsp$ mass is much worse, around 50\%, with the 
central value $1.8\sigma$ from the true one. 
This is due to its small mass, when compared
to the other mass scales, which renders the $m_{T2}^{max}$ expressions weakly dependent on $m_{\tnu}$. Nonetheless,
the results clearly point to a very light LSP, with a mass scale much smaller than the other particles involved in
the cascade decay.

From the results in Fig.~\ref{fig:mt2max} and Table~\ref{table:masses} we conclude that, despite the lack of precision
in determining the LSP mass, the $m_{T2}$ subsystem method can still show that the LSP state is much lighter than
expected in most MSSM scenarios. This would provide strong evidence for a sneutrino DM scenario, at least for the
case of a light gluino/heavy squark spectrum, such as in the SN1 point.

\begin{table}[t]
\centering
\begin{tabular}{|l|c|c|c|c|c|}
\hline
 & $m_{\tg}$ & $m_{\tq_L}$ & $m_{\tw_1}$ & $m_{\tz_1}$ & $m_{\tnu}$  \\
\hline
SN1 & $771\pm 9$ & -- & $236\pm 11$ & -- &$34 \pm 15$ \\
Exact value & 765 & 1520--1523 & 227 & 109 & 7.6 \\
\hline
SN2 & --& $786\pm 4$ & $207\pm 10$ & $126\pm 13$ & $14\pm 8$ \\
Exact value & 765 & 775--779 & 228  & 109 & 7.6 \\
\hline
SN3 & -- & $710 \pm 2$ & $222 \pm 5$ & $111\pm 7$ & $0 \pm 12$ \\
Exact value & 1000 & 700--704 & 228 & 109 & 7.6   \\
\hline
\end{tabular}
\caption{Measured mass values in GeV for the points SN1--SN3, obtained using the $m_{T2}$ subsystem
method described in the text. The error shown only includes statistical uncertainties and assumes
an integrated luminosity of 100~fb$^{-1}$.}
\label{table:masses}
\end{table}

\subsection{Results for SN2 and SN3}
\label{mt2results2}

For points SN2 and SN3 we must consider the squark cascade decay shown in Eq.~(\ref{eq:evtop2}).
To select $\tq_L \to \tw_1 + q$ events we use the following set of cuts:
\begin{equation}
  \begin{array}{l}
    \eslt > 100\;{\rm GeV},~ n(j) = 2,~ n(l) = 2,~ n(b) = 0,\\ 
    p_T(j_1) > 100\;{\rm GeV},~ p_T(j) > 30\;{\rm GeV},~ p_T(l) > 30\;{\rm GeV} 
  \end{array} \label{eq:cuts2}
\end{equation}
and once again veto OS-dilepton events with $80\;{\rm GeV} < m(l^+l^-) < 100$~GeV or $m(l^+l^-) < 40$~GeV.
The cross-section after cuts for points SN2 and SN3 are 32~fb and 36~fb, respectively, while for the SM background we obtain 29~fb.
Despite the large background, the bulk of the SM events are concentraded at low $m_{T2}$ values ($< 500$~GeV). Thus
the position of the $m_{T2}$ end point extraction is almost unaffected by the SM BG, as seen in Fig.~\ref{fig:mt22}.
For point SN2 only $\sim 65\%$ of the signal comes from squark pair production, due to
contamination from gluino pair production and gluino-squark production. On the other hand, the
SUSY background is much smaller for point SN3, with $\sim 82\%$ of the signal coming from squark pair production.
After repeating the same procedure used for extracting the masses in the SN1 case, but using the appropriate $m_{T2}^{max}$ expressions
for the squark decay chain, Eqs.~(\ref{eq:mt2ql}) and (\ref{eq:mt2q}), we obtain the best fit result for $m_{\tq}$, $m_{\tw_1}$ and $m_{\tnu}$
shown in Table~\ref{table:masses}. As we can see, for both points the statistical error bars are smaller than the ones for point SN1, due to the
larger signal cross-section. Once again the least precise measurement corresponds to $m_{\tnu}$, due to its small value. Nonetheless
we can still conclude that the LSP is much lighter than the chargino and squark.

\FIGURE[t]{
\includegraphics[width=10cm]{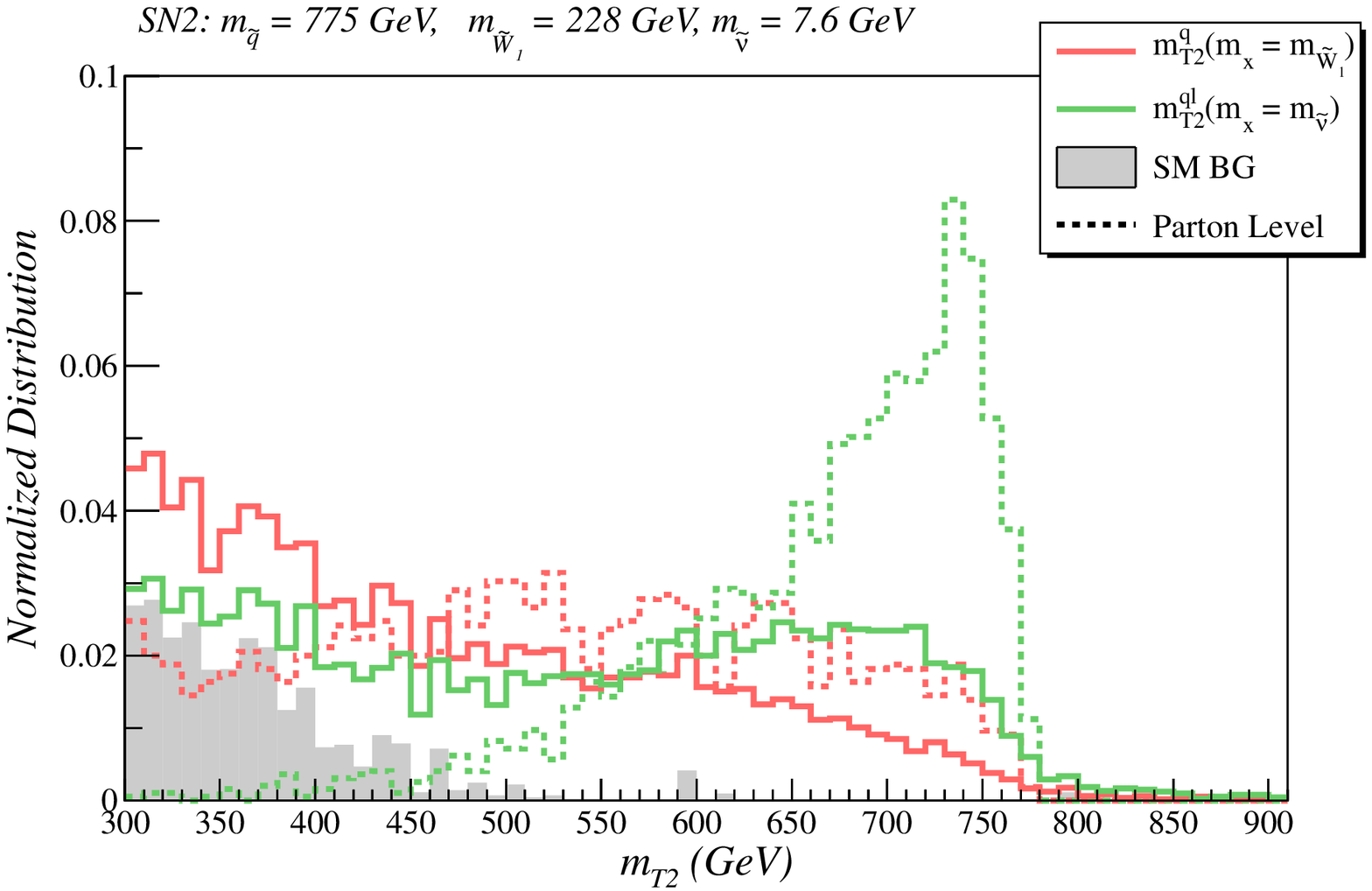}
\includegraphics[width=10cm]{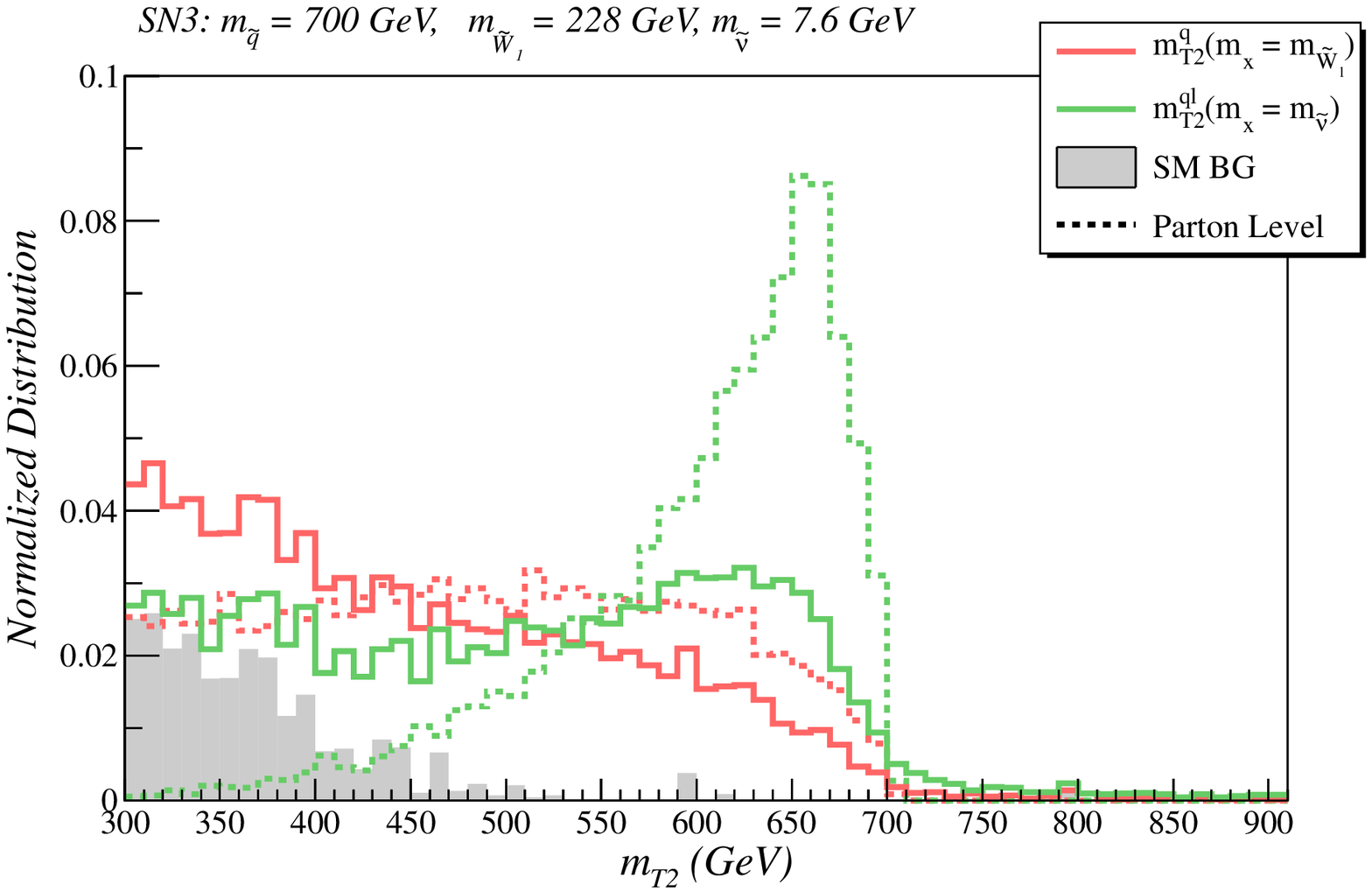}
\caption{The $m_{T2}^{q}$ and $m_{T2}^{ql}$ distributions for the points SN2 and SN3.
The dashed lines show the signal distributions at parton level, the shaded histogram shows the SM contribution and
the solid lines show the signal plus background distributions at the detector level,
after the cuts Eq.~(\ref{eq:cuts2}) have been applied. 
All distributions are normalized to unity.
}\label{fig:mt22}}

As mentioned before, the large $\tq\tq$ production cross-section for points SN2 and SN3 can still provide 
one more piece of information about the spectrum. From Table~\ref{tab:decays},
we have BR$(\tq_R \to q + \tz_1)\sim 100\%$. Therefore, if instead of $\tq_L$ pair production
we consider the $\tq_R\tq_R$ events, we can use the usual $m_{T2}$ variable with $\tq_R$ as the
parent, $\tz_1(\nu\tnu)$ as the daughter and the jet as the visible component to measure
the $\tz_1$ mass once $m_{\tq_R}$ is known. 
In the scenarios considered we have $m_{\tq_L} \simeq m_{\tq_R}$, so we
can use the $m_{\tq}$ value obtained from our previous results.
To select the right-handed squark signal we require:
\begin{equation}
    \eslt > 200\;{\rm GeV},~ n(j) = 2,~ n(l) = 0,~ n(b) = 0,~ p_T(j_1) > 100\;{\rm GeV} \label{eq:cuts3b}
\end{equation}
The SN2 (SN3) signal in this channel is 133 (176)~fb, while for the SM BG we have 68~fb. Once again, the
distribution for the SM background peaks at low $m_{T2}$ and has almost no impact on the $m_{T2}^{max}$ value.
In Fig.~\ref{fig:mt23b} we show the $m_{T2}$ distribution for SN3 with $m_{x} = m_{\tz_1}$, where
we see a clear edge at $m_{T2} \sim 690$ GeV, very close to the $m_{\tq_R}$ input value.

\FIGURE[t]{
\includegraphics[width=10cm]{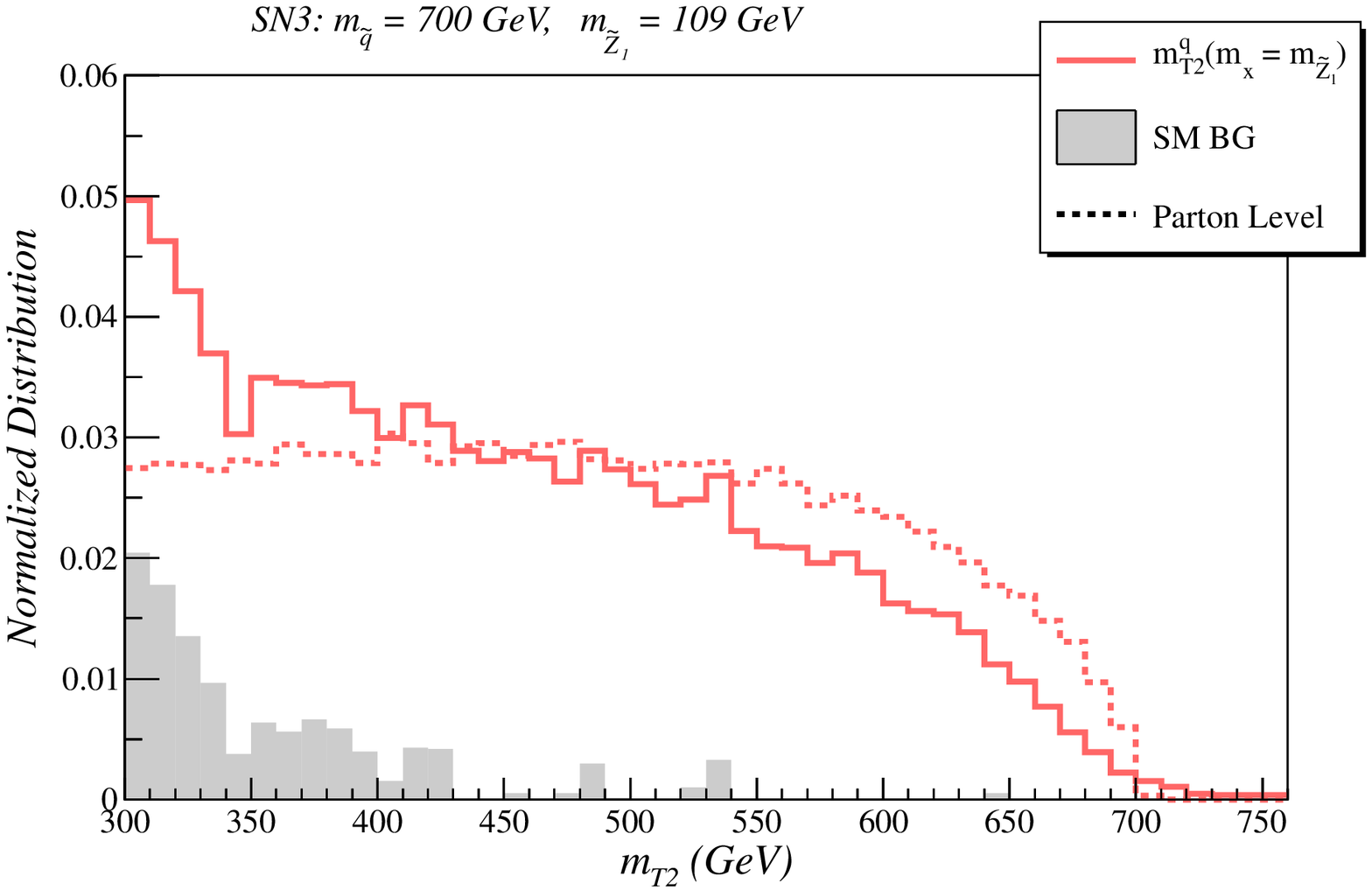}
\caption{The $m_{T2}^{q}$ distribution for point SN3. The dashed line
shows the signal distribution at parton level, while
the solid lines show the signal plus background distribution at the detector level,
after the $\eslt > 200$ GeV, $n(j) = 2$, $n(l) = 0$, $n(b) = 0$, $p_T(j_1) > 100$ GeV
cuts have been applied. All distributions are normalized to 1.
}\label{fig:mt23b}}

The $m_{T2}^{q,max}$ values as a function of $m_x$ for the SN3 point are shown in Fig.~\ref{fig:mt2max3b}.
We see that the extracted endpoints are very close to their exact value and due to the large signal statistics and low BG,
the error bars are barely visible. Fitting Eq.~(\ref{eq:mt2q}) to the data points in Fig.~\ref{fig:mt2max3b} we obtain:
\be
\frac{m_{\tq_R}^2 - m_{\tz_1}^2}{2m_{\tq_R}} =  346 {\rm GeV}
\ee
which agrees extremely well with the theoretical value, 342 GeV. Assuming $m_{\tq_L}=m_{\tq_R}$
and using the $m_{\tq_L}$ value obtained from the $\tq_L\tq_L$ signal (see Table~\ref{table:masses}), we can compute $m_{\tz_1}$:
\be
m_{\tz_1} = 111 \pm 7\; {\rm GeV\; (SN3)}
\ee
where the error in $m_{\tq_R}$ has been included when computing the uncertainty on $m_{\tz_1}$.
Repeating the same procedure for point SN2 we obtain:
\be
m_{\tz_1} = 126 \pm 13\; {\rm GeV\; (SN2)}
\ee
The result in this case is worse than for point SN3 because of the larger SUSY background present in the SN2 signal. This mainly
affects the determination of $m_{\tq}$, as seen in Table~\ref{table:masses}, which propagates to the central value and uncertainty in
$m_{\tz_1}$.  Nonetheless,  the $\tq_R\tq_R$ channel still shows that the measured spectrum has a neutral NLSP with
mass $\approx m_{\tw_1}/2$. Such a neutral NLSP is consistent with a bino state, if gaugino mass unification
is assumed. Therefore, the combined results of the $\tq_L\tq_L$ and $\tq_R\tq_R$ channels would point to
the usual MSSM scenario with gaugino mass unification, but with an additional LSP state, which is neutral
and very light. This would be another strong evidence for the SNDM model.

\FIGURE[t]{
\includegraphics[width=10cm]{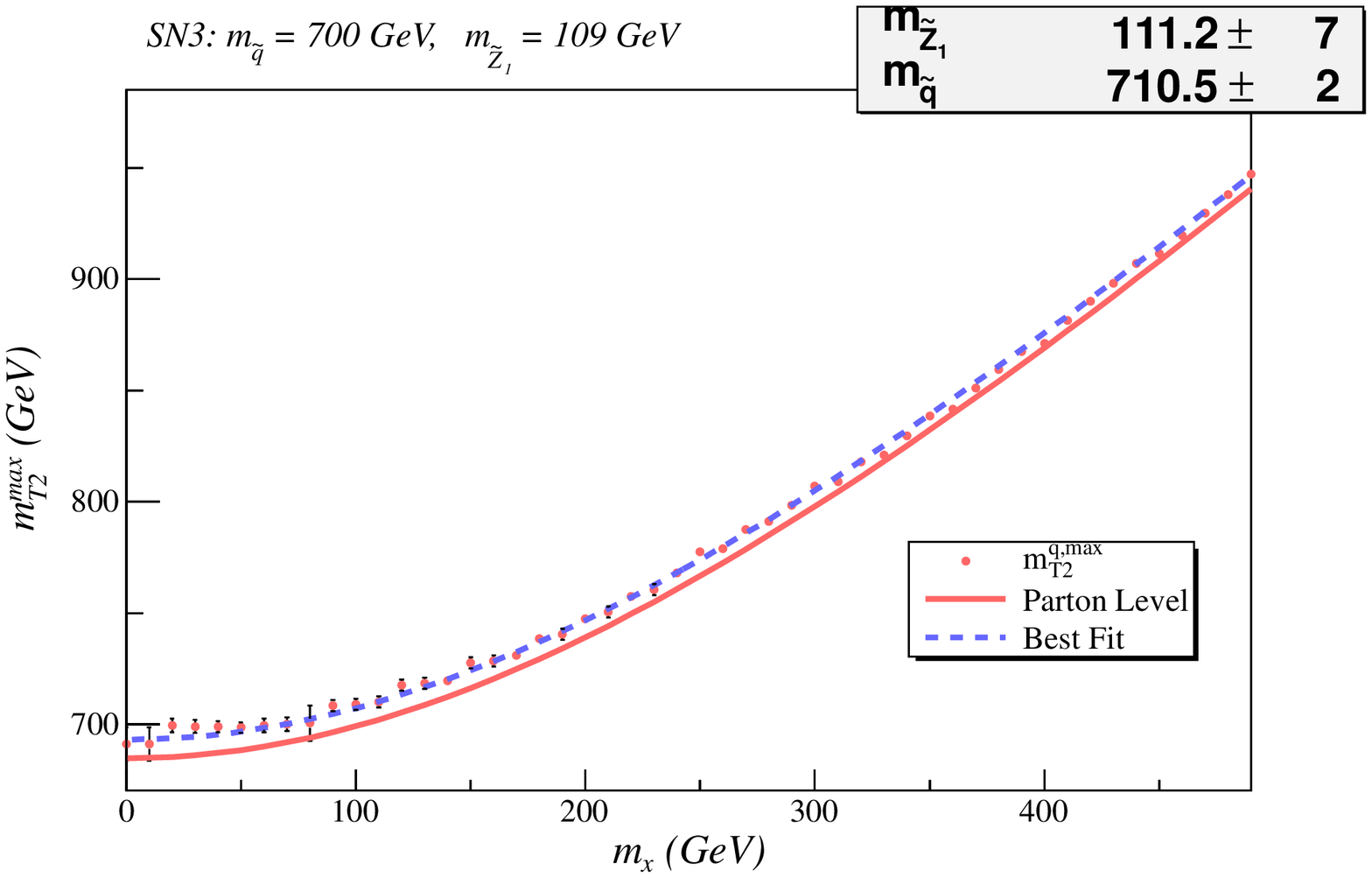}
\caption{$m_{T2}^{max,q}$ for point SN3, as a function of the trial daughter mass $m_x$, in the zero lepton, dijet
channel, as discussed in the text.  The dashed blue line shows the best fit result obtained
using Eq.~(\ref{eq:mt2q}). The solid line shows the exact result obtained at parton level.
}\label{fig:mt2max3b}}

\subsection{Dilepton Invariant Mass at LHC14}

From the results in Table~\ref{table:masses} we see that the $m_{T2}$ method can
indicate the presence of a very light LSP neutral particle in the signal.
Moreover, in case of SN2--3, the $m_{T2}^q$ distribution can give information on an 
additional invisible sparticle, consistent with the $\tz_1$ in case of universal gaugino masses.
Additional evidence for a sneutrino LSP can be obtained from the properties of dilepton events, 
as already discussed in Sec.~\ref{sec:lhc7dists}.
Although at 7 TeV the dilepton signal is likely too small to allow for the use of the dilepton 
invariant-mass distributions, at LHC14 these may be exploited to probe the nature of the LSP.

\FIGURE[t]{
\includegraphics[width=10cm]{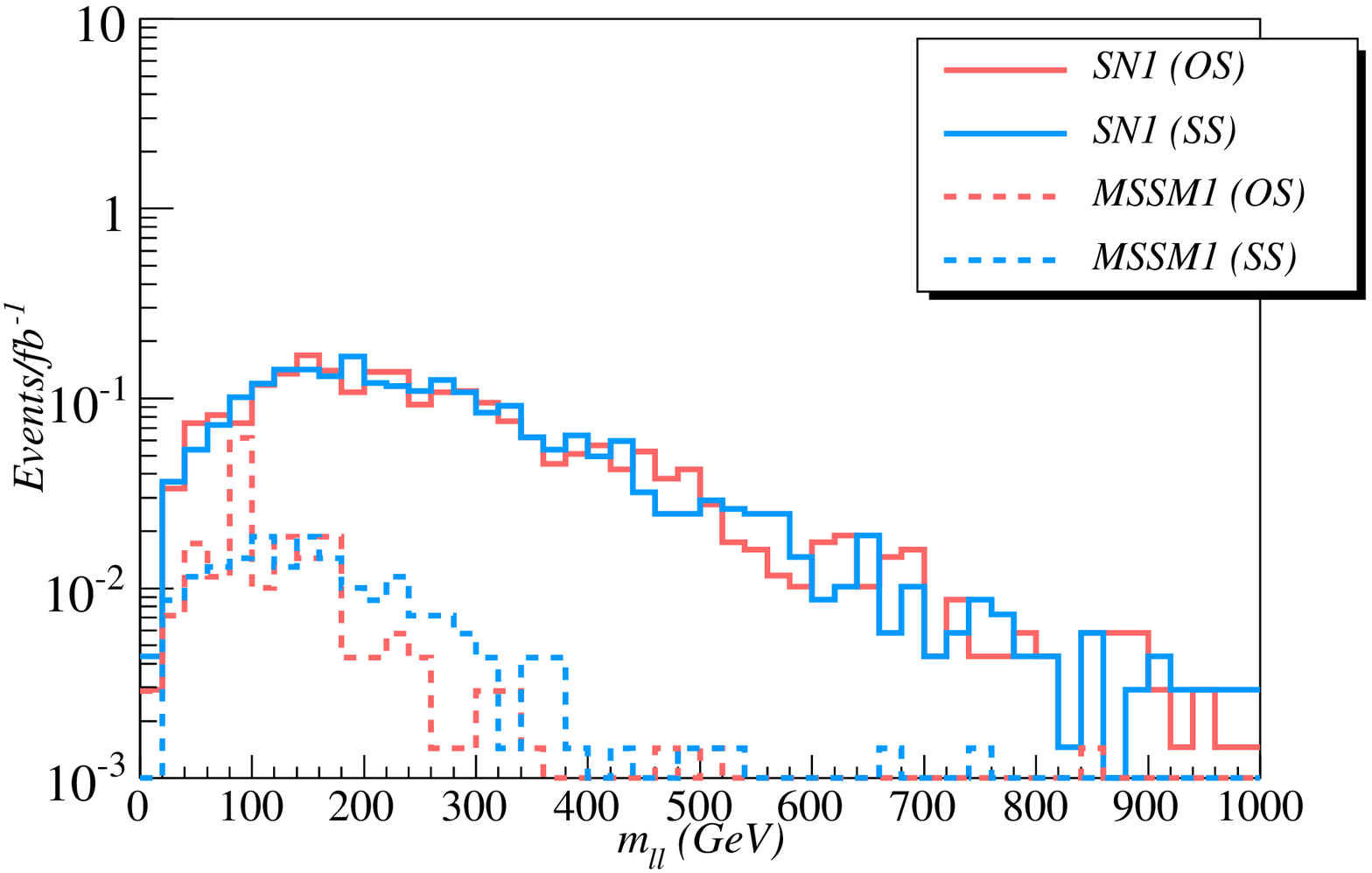}
\includegraphics[width=10cm]{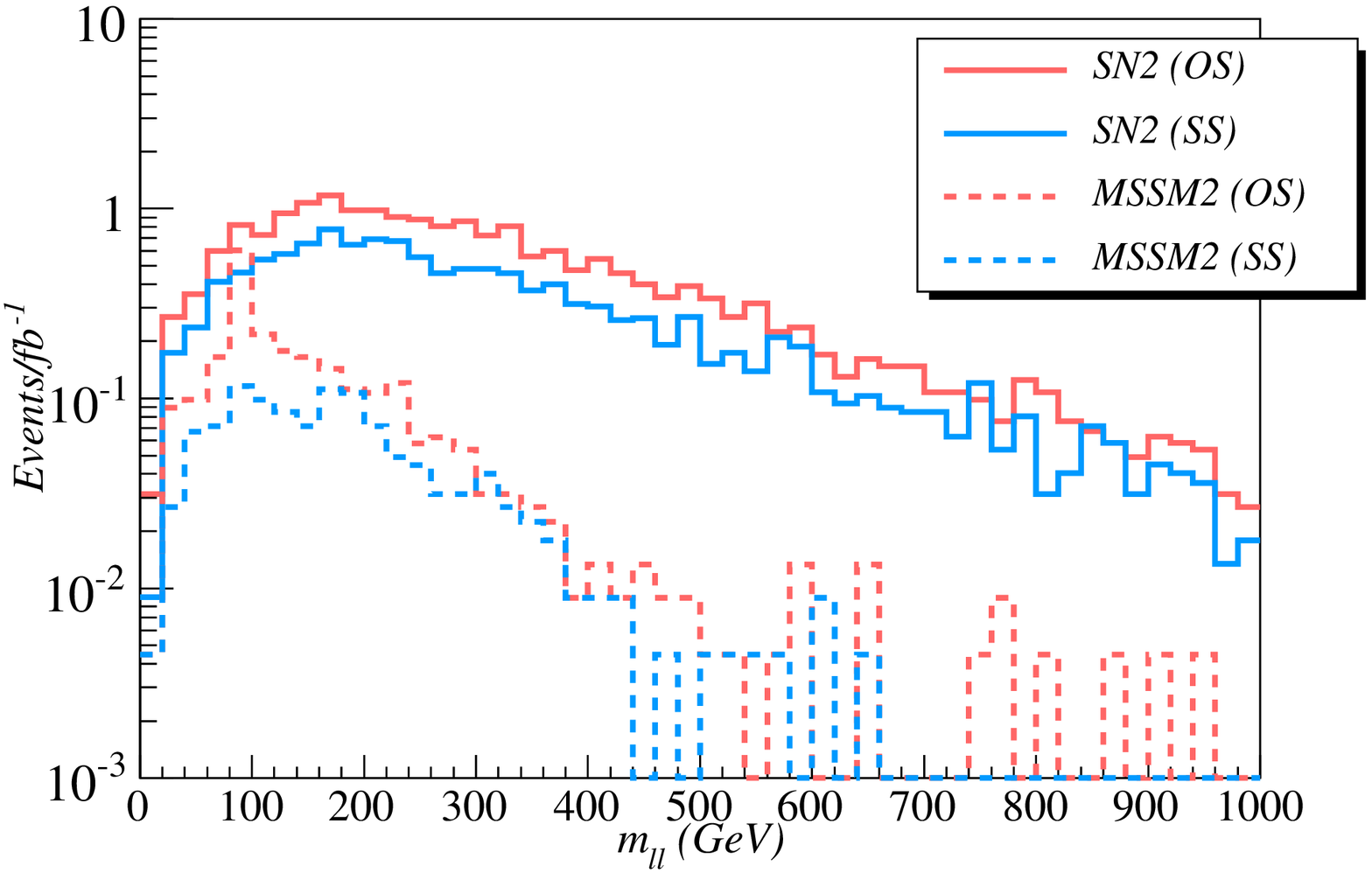}
\includegraphics[width=10cm]{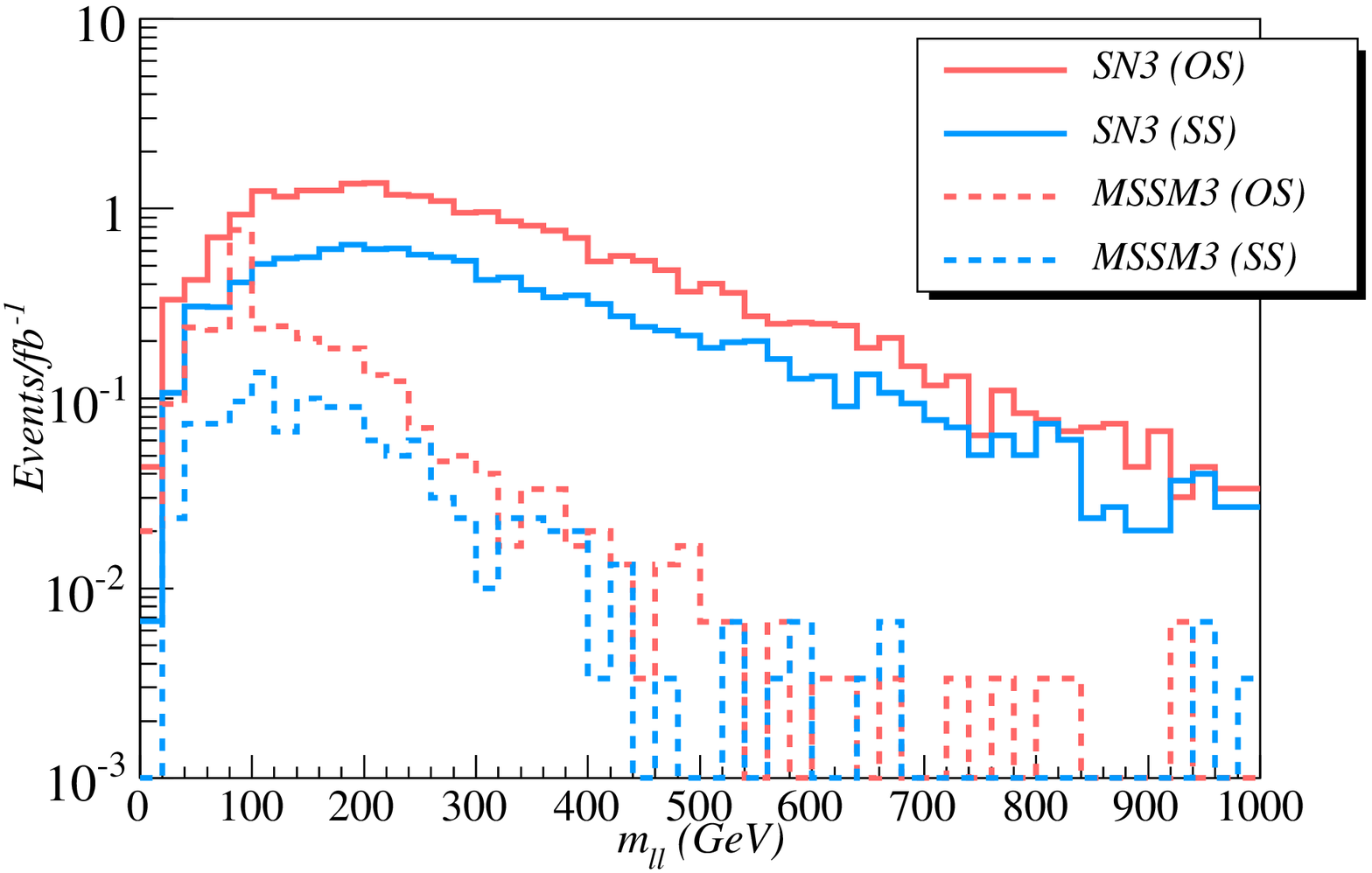}
\caption{OS (red) and SS (blue) dilepton invariant masses for the SN1-3 points (solid) and
the corresponding MSSM models (dashed). For frame {\it a)} the cuts in Eq.(\ref{eq:cuts1})
have been applied, while for frames {\it b)}  and {\it c)} the cuts in Eq.(\ref{eq:cuts2}) were used instead.}\label{fig:dilepmass}
}

In Fig.~\ref{fig:dilepmass} we show the OS and SS dilepton invariant masses for the three SNDM benchmark points as well as for the corresponding MSSM models. The cuts applied are the 
same used for the $m_{T2}$ analyses, namely Eq.~(\ref{eq:cuts1}) for point SN1 and Eq.~(\ref{eq:cuts2}) for points SN2--3. We assume that the shape of the SM background distributions can be 
extracted from data and/or MC, so we neglect their contribution. 
As we can see, the $m_{l^{+}l^{-}}$ and $m_{l^{\pm}l^{\pm}}$ distributions 
are drastically different between the SNDM and MSSM scenarios. 
Besides the larger overall rate of dilepton events, the SNDM points show a much harder 
$m_{ll}$ distribution. This is due to the large $\tw_1-\tnu$ mass gap, resulting in a much 
harder $p_T(l)$ spectrum. Moreover, for the gaugino masses considered here,
the $\tz_2$ has a significant BR to $Z+\tz_1$, giving rise to the Z-peak seen in the MSSM 
OS distributions; in the SNDM scenario, no such peak is present since 
BR$(\tz_2 \to \tnu + \nu) \simeq 100\%$.

\section{Conclusions}
\label{sec:conclude}

A mainly right-handed sneutrino with large L/R mixing is an excellent candidate for 
light cold dark matter with mass below $\sim 10$~GeV.
If DM is indeed realized in the form of light sneutrinos,
there are important consequences to the SUSY signatures at the LHC.
In particular, neutralinos $\tz_1$ and $\tz_2$ appearing in squark and gluino cascades 
decay invisibly into  $\lsp\nu$, so that there can be up to three different invisible 
sparticles in an event. Charginos, on the other hand, decay dominantly into charged 
leptons plus the $\lsp$ LSP.
SUSY events will therefore present a harder $\eslt$ distribution than expected
in MSSM scenarios with a similar sparticle spectrum, and dilepton
events will appear at much larger rates both in the OS and SS channels.
During the first LHC run, at $\sqrt{s}=$7 TeV and with $\sim 1$~fb$^{-1}$,
a signal could already be seen if gluinos and squarks have masses up to $\sim 1$~TeV.
Signal distributions such as the lepton and jet number, as well as SS/OS dilepton rates 
may already indicate a light sneutrino as the lightest SUSY particle in the early phase 
of LHC running.

Precision measurements enabling a model discrimination should be possible at higher 
energy and luminosity. 
For $\sqrt{s}=$14 TeV and $\mathcal{L} = 100$~fb$^{-1}$ we have shown
that the sneutrino mass can be measured using the $m_{T2}$ technique with a $\sim 50\%$
precision, which is already sufficient to distinguish between the SNDM and MSSM scenarios with gaugino mass unification.
Furthermore, the dilepton invariant mass distributions can also point to the presence
of a light LSP which carries lepton number. 
The presence of additional invisible sparticles in the decay chains may be inferred 
from the $\eslt$ and transverse-mass distributions. We have shown that indeed, for 
$m_{\tilde q_R}^{}\approx m_{\tilde q_L}^{}$ the $\tz_1$ mass might be measurable 
with $\sim10\%$ precision. 

Regarding alternative scenarios with possibly similar signatures, a 7--8~ GeV $\tz_1$ LSP in the 
MSSM with non-universal gaugino masses~\cite{Fornengo:2010mk} (see however \cite{Vasquez:2010ru})
could be distinguished from the case studied here by exploiting, e.g., same-flavor opposite-sign 
(SFOS) dileptons from $\tz_2\to \tz_1+Z$, which is absent in the SNDM case. 
Indeed the absence of kinematical structure and 
flavor correlations typical for the SNDM case will point to $\tw_1\to l^\pm\lsp$ decays.\footnote{In case 
of a strong hierarchy among the $\lsp$ of different flavors there can of course be SFOS 
dileptons from cascades involving $\tw_1$'s; however in that case there should also appear 
the corresponding SFSS events.}  
Quite similar signals as in the SNDM case can in principle arise in the next-to-MSSM, 
with $<10$~GeV neutralinos as viable DM candidates that can have 
large ($10^{-5}-10^{-4}$~pb) elastic
scattering cross sections~\cite{Vasquez:2010ru,Draper:2010ew,Cao:2011re}.
In this case one may have dominantly invisible $\tz_2$ decays through  $\tz_2\to h_2\tz_1$
followed by $h_2\to\tz_1\tz_1$. Here possible ways of discrimination are, e.g., 
the $\tw_1\to W^\pm\tz_1$ decays and the presence of additional 
light Higgs states $h_1$ and $a_1$. 

We conclude that the LHC offers very good prospects to resolve the light mixed sneutrino DM case. 
Finally, recall that a corroborating signal is expected in direct dark matter searches, as there is a lower 
limit on the spin-independent scattering cross section of $\sigma^{\rm SI}\gtrsim10^{-5}$~pb.

\acknowledgments

GB and AL thank the LPSC Grenoble for hospitality. 
AL would like to thank Xerxes Tata for useful discussions, 
and SK gratefully acknowledges discussions with Sanjay Padhi on 
dilepton searches in CMS. 
Last but not least, we thank A.~Pukhov for providing 
SHLA decay-table output in {\tt micrOMEGAs}.  

This research was supported in part by the U.S. Department of Energy,
by the Fulbright Program and CAPES (Brazilian Federal Agency for
Post-Graduate Education), and by the French 
ANR project {\tt ToolsDMColl}, BLAN07-2-194882.

\appendix

\section{Endpoint Extraction Algorithm}
\label{sec:extr}

Here we describe the algorithm implemented to extract the 
edge from the $m_{T2}$ distributions. As shown by the solid histograms
 in Figs.\ref{fig:mt2}, \ref{fig:mt22} and \ref{fig:mt23b}, after the inclusion
of both SUSY and SM backgrounds as well as ISR, FSR and detector effects,
the $m_{T2}$ distribution presents a tail beyond the expected $m_{T2}^{max}$
edge. In order to obtain $m_{T2}^{max}$ from the $m_{T2}$ distribution
we use a simple linear {\it kink} model to fit a subset of the $m_{T2}$ bins.
Although the  linear {\it kink} fit has been used in several $m_{T2}$ studies,
we have found that in general the result is dependent on the region
to which the fit is applied. In order to have an unbiased selection
of the region considered in the fit, we apply the following procedure:
\begin{enumerate}
\item First we select a region consisting of all bins after the peak of the $m_{T2}$ distribution,
such that the initial bin in this region ($M_I$) is the peak location and the final bin ($M_F$) is the highest $m_{T2}$ value.
\item Then we perform a fit to the $m_{T2}$ distribution in the $(M_I,M_F)$ interval, using the linear {\it kink} fucntion:
\begin{equation}
f(m_{T2}) = \left\{ 
\begin{array}{lr}
a m_{T2} + b &\mbox{, if $m_{T2} < M$}\\
c (m_{T2} - M) + aM + b &\mbox{, if $m_{T2} > M$}
\end{array} \right. \label{eq:linfit}
\end{equation}
where $a,\;b,\;c$ and $M$ are the free parameters to be fitted. The best fit
value for $M$ provides an estimative for $m_{T2}^{max}$.
\item We then select a new region around the best fit $M$ value, consisting of the interval
$(M_I,M_F)$, where $M_I$ is the same value used for the fit in step 2 minus one bin and $M_F = 2(M - M_I) + M$.
\item We then repeat steps 2 and 3 until $(M_I,M) <$ 5 bins.
\item For each of the intervals $(M_I,M_F)$ we compute the chi-square/degrees of freedom ($\chi^2/ndf$)
for the best fit using Eq.~(\ref{eq:linfit}) and select the $(M_I,M_F)$ interval which gives the lowest $\chi^2/ndf$ value.

\item Finally we fit the $m_{T2}$ distribution in the selected $(M_I,M_F)$ interval and take $m_{T2}^{max}$
as the best fit $M$ value. The error on $m_{T2}^{max}$ is then computed from $\chi^2 -\chi^2_{min} = 1$,
marginalized over the $a$, $b$ and $c$ parameters.
\end{enumerate}
The main idea behind the above procedure is to test distinct $m_{T2}$ intervals and select the one
that is better described by Eq.~(\ref{eq:linfit}). Therefore, it
provides a general way of selecting the proper region to be fitted by the linear {\it kink} function.

Although more sophisticated methods will eventually be used in real
data analysis, the procedure we have implemented is sufficient
for our purposes. Furthermore, we have verified that the procedure described here
works for different SUSY models and $m_{T2}$ subsystems, despite the distinct shapes
of the $m_{T2}$ distributions.

	
%
\bibliographystyle{h-physrev}
\bibliography{sndm5}

\begin{thebibliography}{10}

\bibitem{Bertone:2010zz}
G.~Bertone, editor,
\newblock {\em {Particle dark matter: Observations, models and searches}}
  (Cambridge University Press, 2010).

\bibitem{Jungman:1995df}
G.~Jungman, M.~Kamionkowski, and K.~Griest,
\newblock Phys.Rept. {\bf 267}, 195 (1996), hep-ph/9506380.

\bibitem{Baer:2008uu}
H.~Baer and X.~Tata,
\newblock (2008), 0805.1905,
\newblock and references therein.

\bibitem{Bilenky:1998dt}
S.~M. Bilenky, C.~Giunti, and W.~Grimus,
\newblock Prog.Part.Nucl.Phys. {\bf 43}, 1 (1999), hep-ph/9812360.

\bibitem{ArkaniHamed:2000bq}
N.~Arkani-Hamed, L.~J. Hall, H.~Murayama, D.~Tucker-Smith, and N.~Weiner,
\newblock Phys.Rev. {\bf D64}, 115011 (2001), hep-ph/0006312.

\bibitem{Borzumati:2000mc}
F.~Borzumati and Y.~Nomura,
\newblock Phys. Rev. {\bf D64}, 053005 (2001), hep-ph/0007018.

\bibitem{Komatsu:2010fb}
WMAP Collaboration, E.~Komatsu {\em et~al.},
\newblock Astrophys.J.Suppl. {\bf 192}, 18 (2011), 1001.4538.

\bibitem{Jarosik:2010iu}
N.~Jarosik {\em et~al.},
\newblock Astrophys.J.Suppl. {\bf 192}, 14 (2011), 1001.4744.

\bibitem{Belanger:2010cd}
G.~Belanger, M.~Kakizaki, E.~Park, S.~Kraml, and A.~Pukhov,
\newblock JCAP {\bf 1011}, 017 (2010), 1008.0580.

\bibitem{Mt2sub}
M.~Burns, K.~Kong, K.~T. Matchev, and M.~Park,
\newblock JHEP {\bf 03}, 143 (2009), 0810.5576.

\bibitem{Thomas:2007bu}
Z.~Thomas, D.~Tucker-Smith, and N.~Weiner,
\newblock Phys.Rev. {\bf D77}, 115015 (2008), 0712.4146.

\bibitem{Suspect}
A.~Djouadi, J.-L. Kneur, and G.~Moultaka,
\newblock Comput. Phys. Commun. {\bf 176}, 426 (2007), hep-ph/0211331.

\bibitem{Pythia}
T.~Sjostrand, S.~Mrenna, and P.~Z. Skands,
\newblock JHEP {\bf 05}, 026 (2006), hep-ph/0603175.

\bibitem{Kraml:2007sx}
S.~Kraml and D.~Nhung,
\newblock JHEP {\bf 0802}, 061 (2008), 0712.1986.

\bibitem{Kumar:2009sf}
A.~Kumar, D.~Tucker-Smith, and N.~Weiner,
\newblock JHEP {\bf 09}, 111 (2010), 0910.2475.

\bibitem{MarchRussell:2009aq}
J.~March-Russell, C.~McCabe, and M.~McCullough,
\newblock JHEP {\bf 03}, 108 (2010), 0911.4489.

\bibitem{Alpgen}
M.~L. Mangano, M.~Moretti, F.~Piccinini, R.~Pittau, and A.~D. Polosa,
\newblock JHEP {\bf 07}, 001 (2003), hep-ph/0206293.

\bibitem{Calchep}
A.~Pukhov,
\newblock (2004), hep-ph/0412191.

\bibitem{Belanger:2006is}
G.~Belanger, F.~Boudjema, A.~Pukhov, and A.~Semenov,
\newblock Comput.Phys.Commun. {\bf 176}, 367 (2007), hep-ph/0607059.

\bibitem{Skands:2003cj}
P.~Z. Skands {\em et~al.},
\newblock JHEP {\bf 0407}, 036 (2004), hep-ph/0311123.

\bibitem{isajet}
F.~E. Paige, S.~D. Protopopescu, H.~Baer, and X.~Tata,
\newblock (2003), hep-ph/0312045.

\bibitem{Baer14}
H.~Baer, V.~Barger, A.~Lessa, and X.~Tata,
\newblock JHEP {\bf 09}, 063 (2009), 0907.1922.

\bibitem{Matchev:2009iw}
K.~T. Matchev, F.~Moortgat, L.~Pape, and M.~Park,
\newblock JHEP {\bf 0908}, 104 (2009), 0906.2417.

\bibitem{Barr:2011xt}
A.~Barr {\em et~al.},
\newblock (2011), 1105.2977.

\bibitem{Lester1999}
C.~G. Lester and D.~J. Summers,
\newblock Phys. Lett. {\bf B463}, 99 (1999), hep-ph/9906349.

\bibitem{Barr2003}
A.~Barr, C.~Lester, and P.~Stephens,
\newblock J. Phys. {\bf G29}, 2343 (2003), hep-ph/0304226.

\bibitem{Cho2007}
W.~S. Cho, K.~Choi, Y.~G. Kim, and C.~B. Park,
\newblock JHEP {\bf 02}, 035 (2008), 0711.4526.

\bibitem{Rajaraman2010}
A.~Rajaraman and F.~Yu,
\newblock (2010), 1009.2751.

\bibitem{Fornengo:2010mk}
N.~Fornengo, S.~Scopel, and A.~Bottino,
\newblock Phys. Rev. {\bf D83}, 015001 (2011), 1011.4743.

\bibitem{Vasquez:2010ru}
D.~A. Vasquez, G.~Belanger, C.~Boehm, A.~Pukhov, and J.~Silk,
\newblock Phys. Rev. {\bf D82}, 115027 (2010), 1009.4380.

\bibitem{Draper:2010ew}
P.~Draper, T.~Liu, C.~E. Wagner, L.-T. Wang, and H.~Zhang,
\newblock Phys.Rev.Lett. {\bf 106}, 121805 (2011), 1009.3963.

\bibitem{Cao:2011re}
J.~Cao, K.-i. Hikasa, W.~Wang, and J.~M. Yang,
\newblock (2011), 1104.1754.

\end{thebibliography}

\end{document}